\documentclass[a4paper,10pt]{article}
\usepackage[utf8]{inputenc}
\usepackage{xspace}		
\usepackage[english]{babel}
\usepackage[T1]{fontenc}
\usepackage{lmodern}
\usepackage{authblk}

\usepackage[margin=25mm,bindingoffset=10mm]{geometry} 
\usepackage{setspace}
\usepackage{ragged2e}
\usepackage[all]{xy}
\usepackage{graphicx} 
\usepackage{color}
\usepackage[svgnames]{xcolor}

\usepackage{pdfpages}
\usepackage{amsmath, mathtools}
\usepackage{amssymb,amsfonts,dsfont}
\usepackage{mathrsfs}
\usepackage{amsthm}
\usepackage{bm}				
\usepackage{bbm,upgreek}
\usepackage{makecell}
\usepackage{cellspace}
\usepackage{hyperref} 
\usepackage{cleveref}
\hypersetup{linktocpage,colorlinks,citecolor={blue},pdfdisplaydoctitle=true,pdfpagemode=UseOutlines,bookmarksnumbered=true}
\usepackage[sort&compress,numbers]{natbib}
\usepackage{url}
\usepackage{doi}

\newcommand{\bra}[1]{\langle #1 |} 
\newcommand{\ket}[1]{| #1 \rangle } 
\definecolor{cbl}{rgb}{0,0,1}
 
\definecolor{crd}{rgb}{1,0,0}
 
\newcommand{\upd}{\mathrm{d}}
\newcommand{\tr}{\mathrm{tr}}

\newcommand{\ie}[0]{\textit{i.e.}}
\newcommand{\eg}[0]{\textit{e.g.}}

\newcommand\e{\mathrm{e}}

\definecolor{nblue}{rgb}{0.06,0.3,0.73}
\definecolor{nblack}{rgb}{0,0,0}
\definecolor{nred}{rgb}{0.9,0.1,0.1}

\DeclareMathOperator*{\argmin}{argmin}

\title{A study of the quantum Sinh-Gordon model with relativistic continuous matrix product states}

\author{Antoine Tilloy\footnote{\url{antoine.tilloy@gmail.com}} }

\affil{\vskip-0.2cm \it \small  LPENS, Département de physique, École Normale Supérieure - PSL

Centre Automatique et Systèmes (CAS), Mines Paris - PSL

Université PSL, Sorbonne Université, CNRS, Inria, 75005 Paris

 \normalsize}

\date{}

\begin{document}

\maketitle

\begin{abstract}
\noindent I study the Sine-Gordon (SG) and Sinh-Gordon (ShG) quantum field theories with a recently introduced variational method, the relativistic continuous matrix product states (RCMPS). The main advantage is to work directly in the thermodynamic limit, and without any UV regulator. The SG model is well understood and integrable, which provides a convenient benchmark for the variational method and serves as a warm-up. RCMPS approximate the ground state of the SG model arbitrary well up to the free Fermion point [coupling $\beta=\sqrt{4\pi}$ in equal-time quantization convention, or $b=1/\sqrt{2}$ in CFT convention], where the ground energy collapses to $-\infty$, and some renormalized ansatz would be needed. The ShG model, while integrable, is less understood and its strong coupling regime $\beta \approx 1$ is subject to some controversy. RCMPS also fit the ground state of the ShG model up to approximately $b=1/\sqrt{2}$, after which their predictions start to deviate substantially from the ``exact'' results. This is more puzzling as nothing is expected to happen physically for the ShG model at that point (\eg\, the ground energy density does not diverge). Either the ``exact'' ShG results are not exact (the analytic continuation of the SG Bethe Ansatz solution is unwarranted), or, more likely, the physical structure of the ShG ground state changes in such a way that it becomes out of reach of the RCMPS manifold for reasonable bond dimensions.
\end{abstract}

\section{Introduction}

The Sinh-Gordon (ShG) model is the 1+1 dimensional scalar quantum field theory (QFT) with a $\cosh(\beta\phi)$ potential, \ie\, with Hamiltonian
\begin{equation}
    H = \int \frac{\pi^2}{2} + \frac{(\partial_x\phi)^2}{2} + \frac{m^2}{\beta^2} \, \cosh\left[\beta\, \phi(x)\right]\, ,
\end{equation}
where $\phi$ and $\pi$ obey canonical commutation relations $[\phi(x),\pi(y)]=i\delta(x-y)$.
We will give a more precise definition in the following section, in particular regarding the necessary normal-ordering.

This choice of potential is remarkable because it seems like it should give the simplest interacting  QFT in 1+1 dimensions: it is integrable (so a lot is known about it analytically) and contains a single stable particle without particle production\footnote{In fact, one way to get the $\cosh$ term is simply to start from the self-interacting scalar field theory with $\phi^4$ potential, and add $\phi^{2n}$ terms order by order to cancel particle production terms \cite{mussardo2020}.} (so its physics should be fairly straightforward). However, this simplicity is deceptive, and the domain of validity of the exact formulas for physical quantities of the model is unclear, as some analytic continuations may be unwarranted. In fact, even the range of coupling $\beta$ for which the model is well defined (or even definable) is not fully settled. One salient prediction of integrability is a duality (for certain physical quantities) under the change $\beta \rightarrow 8\pi/\beta$ ($b\rightarrow b^{-1}$ with $b = \beta/\sqrt{8\pi}$), which is surprising as it is not manifest in $H$. Two recent studies, by Konik, L\'ajer, and Mussardo (KLM) \cite{konik2021} and by Bernard and LeClair (BLC) \cite{bernard2022} have seriously put in doubt this duality. In particular, both studies suggest that for $b\geq1$ ($\beta \geq \sqrt{8\pi}$), \ie\, past the self-dual point, the model is massless.

KLM used an impressive range of analytical and numerical methods to approach the self-dual point and test the validity of results from integrability. However, it seems that no matter the approach, the simulation cost explodes precisely as one gets closer to the regime of interest. The main objective of this paper is to try a new variational method, the relativistic continuous matrix product states (RCMPS) \cite{rcmps_letter,rcmps_article}, to push further into the unknown domain. The advantage of this method is to work directly in the continuum and thermodynamic limits. The hope is both to gain understanding about the Sinh-Gordon model itself, and to test RCMPS on a model that is far more difficult to study numerically than $\phi^4$, where RCMPS have been shown to work well. 
\section{The models}

\subsection{Two definitions}
One can define the Sine-Gordon (SG) and Sinh-Gordon (ShG) models from their Hamiltonian formulation, directly in the continuum. As I will later discuss, this definition may be inappropriate for large coupling, when new divergences occur, but let us proceed anyway. To this end, one can first introduce the massless free Boson Hamiltonian (in standard, equal-time quantization):
\begin{equation}
    H_0 = \int \frac{\pi^2}{2} + \frac{(\partial_x\phi)^2}{2}
\end{equation}
and add a (normal-ordered) cosine or hyperbolic cosine potential
\begin{align}
    H_\text{ShG}(\beta) &=\; :\! H_0\! :_m + \int \upd x \; \frac{m^2}{\beta^2} \, :\cosh\left[\beta\, \phi(x)\right]:_m \label{eq:HShG}\\
    H_\text{SG}(\beta) &=\; : \! H_0\! :_m - \int \upd x \; \frac{m^2}{\beta^2} \, :\cos\left[\beta\, \phi(x)\right]:_m \label{eq:HSG}
\end{align}
where $\beta$ is the coupling constant that determines the physics while $m$ simply fixes the scale. Naturally the two models are related and $H_\text{ShG}(\beta) = H_\text{SG}(i\beta)$. The normal-ordering $:\,:_m$ is done with respect to the free creation-annihilation operators of mass $m$, which are related to the field operators by
\begin{align}\label{eq:modeexpansion}
    \phi(x) &= \frac{1}{2\pi} \int \upd k \sqrt{\frac{1}{2 \, \omega_k}} \left(\e^{ikx} a_k + \e^{-ikx} a^\dagger_k \right) \\
        \pi(x) &= \frac{1}{2i\pi} \int \upd k \sqrt{\frac{\omega_k}{2}} \left(\e^{ikx} a_k - \e^{-ikx} a^\dagger_k \right) \, ,
\end{align}
where $\omega_k=\sqrt{k^2 + m^2}$ and $[a_k,a_{k'}^\dagger]=2\pi\delta(k-k')$.
This choice of normal-ordering completely fixes the models while leading to the simplest expressions. Other choices,  with $\tilde{m} \neq m$, lead to a simple change of scale and shift in the vacuum energy.

Both models are sometimes alternatively constructed in radial quantization, which is more natural from the conformal field theory (CFT) perspective. The starting point is the action of the massless free boson
\begin{equation}\label{eq:freeboson_action}
    S_0 = \int \upd^2 z \frac{1}{16\pi} \, (\nabla \varphi)^2(z) \, .
\end{equation}
The interacting models are then obtained on the plane by perturbing the dilation operator $D_0$ of the free boson by vertex operators
\begin{align}
    D_\text{ShG}(b) &= D_0 + \mu_\text{ShG} \int_C\upd z\,  \left[\mathcal{V}_b (z,z^*) + \mathcal{V}_{-b} (z,z^*)\right] \\
    D_\text{SG}(b) &=D_0 - \mu_\text{SG} \int_C\upd z\,  \left[\mathcal{V}_{ib} (z,z^*) + \mathcal{V}_{-ib} (z,z^*)\right]
\end{align}
where $C$ is the unit circle and $\mathcal{V}_b(z,z^*)=:\! \e^{b\varphi(z,z^*)}\! :$ is the vertex operator normal-ordered for the modes of the free field $\varphi$. With the convention taken for the normalization in \eqref{eq:freeboson_action}, the scaling dimension of $\mathcal{V}_b$ is $\Delta=2 b^2$. 

Both constructions are equivalent. Starting from the second definition, mapping the plane to a cylinder of radius $R$, and then taking the radius to infinity, one gets back the first definition \cite{konik2021}, with the normal-ordered vertex operators identified in the following way
\begin{equation}
   :\e^{\sqrt{8\pi} a \phi(x)}\!:_m \; \longleftrightarrow \frac{m^{2a^2}\e^{2a^2 \gamma_E}}{2^{2a^2}  } \; :\e^{a\varphi(x)}\!: \;\;, \label{eq:vertex_equivalence}
\end{equation}
where $\gamma_\text{E}$ is the Euler–Mascheroni constant. The coupling constants in the two definitions thus verify
\begin{align}
    b &= \beta/\sqrt{8\pi}\\
    \mu_\text{ShG} &= \frac{m^{2+2b^2}}{2^{4+2b^2}\pi b^2} \e^{2b^2 \gamma_\text{E}} \label{eq:mushg}\\
    \mu_\text{SG} &=\frac{m^{2-2b^2}}{2^{4-2b^2}\pi b^2} \e^{-2b^2 \gamma_\text{E}} \label{eq:musg}
\end{align}
In addition, one can show (\cite{konik2021}, appendix A) that the ground energy $\varepsilon_0^{(\text{etq})}$ in the equal-time quantization approach is only a constant away from the ground energy $\varepsilon_0^{(\text{rq})}$ in the radial quantization definition 
\begin{equation}\label{eq:energy_dictionary}
    \varepsilon^{(\text{etq})}_0 = \varepsilon^{(\text{rq})}_0 - \frac{m^2}{8\pi} \, .
\end{equation}
This allows to relate the quantities computed from both approaches, which will be particularly convenient for the present study. Indeed, the first equal-time quantization definition is natural to use with the variational method, especially with relativistic continuous matrix product states. This is the one we will use to get numerical results. However, most exact results we will use for comparison have been obtained from the second definition, in radial quantization.

\subsection{Remarkable values of the coupling}
The Hamiltonians \eqref{eq:HShG}-\eqref{eq:HSG} we gave for the Sinh-Gordon and Sine-Gordon models define legitimate QFTs, without the need for any additional renormalization as long as  $b<1/\sqrt{2}$ (equivalently $\beta< \sqrt{4\pi}$) \cite{froehlich1975,froehlich1977}. This is the safe regime, where normal-ordering is provably sufficient to remove all divergences in both models.

For $b\in ]1/\sqrt{2},1[$, the Sine-Gordon model can still be constructed rigorously \cite{dimock1993}, but normal-ordering does not kill all divergences. Informally, the renormalized Hamiltonian is then the same as in \eqref{eq:HSG} up to an infinite counter term proportional to the identity. Without this divergent counterterm, the vacuum energy density of $\eqref{eq:HSG}$ is infinitely negative. This is \emph{a priori} a problem for a variational method, like the one we will explore, that gives finite values of the energy density by construction. For $b>1$ (equivalently $\beta > \sqrt{8\pi}$), the scaling dimension of the cosine potential is larger than $2$ and the interaction is irrelevant. Hence, the Sine-Gordon model is no longer well defined past that point without a short distance cutoff.

The situation is less clear for the Sinh-Gordon model, even though the model \emph{a priori} looks simpler. The scaling dimension $\Delta$ of the $\cosh$ term is always negative, and thus the interaction should (intuitively) always be strongly relevant. However, the model was rigorously constructed by Fr\"ohlich and Park only for $b<1/\sqrt{2}$ \cite{froehlich1977}, and, as far as I know, nothing past that value is established beyond reasonable doubt. The value $\beta=\sqrt{8\pi}$, $b=1$, is remarkable because it corresponds to a \emph{formal} self-dual point ($b\rightarrow 1/b$) of the exact S-matrix. However, it is unclear if this duality is physical when the model is constructed from the definition \eqref{eq:HShG} we provided. In fact, the recent thorough analytical and numerical study of Konik, L\'ajer, and Mussardo (KLM) \cite{konik2021} suggests that the model could be massless for $b >1$. For intermediate values, $b\in ]1/\sqrt{2},1[$, the Hamiltonian truncation (HT) data of KLM is likely not fully converged but still suggests that the Hamiltonian \eqref{eq:HShG} could exist, and that its physical properties could match those predicted by the ``exact'' solution. Quotation marks are warranted because the ``exact'' formulas for the energy density or expectation values of vertex operators are obtained from analytic continuation of formulas derived for the Sine-Gordon model. The validity of the analytic continuation is not in doubt at small coupling, but could break down past a certain threshold. For example, something could \emph{a priori} happen at $b=1/\sqrt{2}$, where the Sine-Gordon model has its first phase transition, or at $b=1$ where there the Sine-Gordon model goes through a BKT transition and seizes to exist without cutoff. KLM seem to favor the second scenario, \ie\, a transition to a massless phase for $b \geq 1$, breaking the self-duality. 

\section{The variational method with RCMPS}

\subsection{Principle of the variational method}
To study the two previously introduced field theories, we will use the variational method so it helps to recall its basic philosophy.
The idea of the variational method is to look for the ground state of the model
by minimization over a carefully chosen submanifold $\mathcal{M}$ of the 
Hilbert space $\mathscr{H}$
\begin{equation}\label{eq:variational}
	\ket{\text{ground}}\simeq \ket{w}\;\;\text{where}\;\;
	\ket{w}=\argmin_{\ket{w}\in\mathcal{M}} \bra{w} \hat{h}\ket{w}\, ,
\end{equation}
where $\hat{h}$ is the Hamiltonian density of the model (assumed to be
translation invariant) and $\ket{w}$ is assumed to be normalized. 

Ideally, the ansatz manifold $\mathcal{M}$ should have 3 
favorable properties: i) computability ii) expressiveness iii) extensivity. 
Computability means that one should be able to compute
efficiently expectation values of local observables for states in $\mathcal{M}$. 
Expressiveness means that one should also
be able to approximate arbitrarily well any state within the Hilbert space
as the dimension of the submanifold is increased (the
ansatz manifold should be dense). Extensivity, finally, means that
the number of parameters (the dimension
of the manifold) should not scale prohibitively with system size for 
a similar level of precision (ideally linearly,
or even independently of system size in the translation invariant case).
In the context of relativistic QFT, an extra constraint is that the ansatz manifold
should be compatible with the singular short distance behavior expected in the 
ground state.

Choosing as ansatz manifold $\mathcal{M}$ a vector space corresponds to
the so called Hamiltonian truncation approach which has successfully been used 
to study a wide range of models including the 
ShG and SG models. Using a vector space unfortunately makes extensivity impossible, 
and thus one needs an IR cutoff to keep a finite number of parameters, but the advantage is that the minimization \eqref{eq:variational} is a trivial quadratic problem.
I will focus instead on an approach that fits the 3 favorable properties, extensivity
included: the relavitivistic continuous matrix product states (RCMPS). This ansatz is a recent extension of the continuous matrix product states (CMPS) of Verstraete and Cirac \cite{verstraete2010}. The minimization 
\eqref{eq:variational} is less trivial but remains feasible using Riemannian optimization techniques. 
The final product is a good 
approximation of the ground state which can be used to compute expectation values
of products of local operators at equal time.
In principle, one can also use the variational method to access excited states and
spectral properties, by diagonalizing the Hamiltonian on the tangent space of
the approximate ground state, but I will not attempt it in this first inquiry.

\subsection{Relativistic continuous matrix product states}
For a bosonic field theory, a relativistic continuous matrix product state 
is an ansatz parameterized by two complex matrices $(Q,R)$ of size $D\times D$
\begin{equation}
    \ket{Q,R} = \tr\left\{\mathcal{P} \exp\left[\int_0^L \upd x \, Q\otimes \mathds{1} 
    + R\otimes a^\dagger(x)\right]\right\}\ket{0}_m.
\end{equation}
In this formula, $\mathcal{P}$ is the path-ordering operator and the trace is taken over the finite matrix space. The creation operator $a^\dagger(x)$ is the standard Fourier transform of the mode creation operator $a^\dagger_k$ (\ie\, without factor $\sqrt{\omega_p}$)
\begin{equation}
    a(x) = \frac{1}{2\pi} \int \upd k \, \e^{ikx} a_k \, .
\end{equation}
This choice may seem awkward from a relativistic QFT perspective, as $a(x)$ does not transform in any nice way under boosts and thus neither does $\ket{Q,R}$. Fortunately, this is not a requirement for a variational ansatz: we just want an ansatz that is computable, maximally expressive, and extensive.

Extensivity of the state is manifest from its exponential form, and in the translation invariant case, the number of parameters is independent from the size of the interval $[0,L]$. Better, without loss of generality (because the parameterization is redundant), one can choose a left canonical gauge $Q=-iK - \frac{1}{2} R^\dagger R$ where $K$ is self-adjoint. With this gauge fixing, the norm of $\ket{Q,R}$ is $1$ in the thermodynamic limit and one can formally define the state on $\mathbb{R}$ directly by its correlation functions. Expressiveness is expected from the fact that, with $x$-dependent diagonal $Q$ and $R$, one can reproduce all sums of field coherent states, which are dense in the Fock space\footnote{The weakness of this argument is that it does not show that one can approximate any translation invariant state in the Fock space with a translation invariant RCMPS (with $Q,R$ independent of $x$). To my knowledge, it is expected to hold but has not been proved.}.
Expectation values of normal-ordered field polynomials and vertex operators on $\ket{Q,R}$ can be computed to machine precision at a cost $\propto D^3$ in the thermodynamic limit (see appendix). Other correlation functions (\eg\, $2$ or $3$ point correlation functions of field monomials or vertex operators) are also computable in principle at a polynomial cost in $D$. Finally, the ansatz reproduces the short distance behavior of the free boson QFT already for $D=0$, since $\ket{R=0,Q=0} = \ket{0}_m$. At least for super-renormalizable models, this short distance behavior is not modified by turning an interaction on, and thus the singular UV behavior should remain well fitted by RCMPS.


\subsection{Optimization}
To find the ground state, one simply tunes the coefficients of the $R,Q$ matrices (or rather $R,K$ in left-canonical gauge) to minimize the Hamiltonian density. The technical details are provided in the appendix \ref{app:geometric}p, and I only sketch the broad picture of how it is done here.

The first step is to compute the gradient of the Hamiltonian density: this can be done with backward differentiation methods, giving the full gradient at a cost only $\propto D^3$.

Naive gradient descent would be extremely inefficient: one would get stuck in plateaus even for fairly moderate bond dimensions. This is because the ``naive'' metric on the matrices parameterizing the RCMPS is very singular with respect to the ``natural'' metric induced on the manifold by the Hilbert space scalar product. Intuitively, some small moves in the $R,K$ matrices change the state a lot, while others have next to no impact. Fortunately, the proper induced metric is efficiently computable, and thus one can do Riemannian gradient descent on the natural manifold. This is actually equivalent to the time dependent variational principle (TDVP) in imaginary time, \ie\, the imaginary time evolution projected onto the RCMPS manifold (which gives another intuition for the efficiency of the approach).

Even better, one can use improved optimization techniques generalized to the Riemannian context, like conjugate gradient or quasi Newton methods such as the limited-memory Broyden-Fletcher-Goldfarb-Shanno (LBFGS) algorithm. They have already been introduced and implemented in the (non-relativistic) MPS \cite{hauru2021} and CMPS \cite{tuybens2022} contexts, and their extension to RCMPS is straightforward.

Finally, one can slightly reduce the dimension of the RCMPS manifold by implementing exactly the $\mathbb{Z}_2$ symmetry present in the ground states of the SG and ShG models, \ie\, the invariance under $\phi \rightarrow -\phi$ which is equivalent to $a^\dagger(x) \rightarrow - a^\dagger(x)$. This can be enforced at the RCMPS level by restricting the ansatz to block matrices
\begin{equation}
    R  = \left(\begin{array}{cc}
       0  & R_1 \\
       R_2  & 0 
    \end{array}\right) ~~ \text{and} ~~
    K  = \left(\begin{array}{cc}
       K_1 & 0 \\
       0  & K_2
    \end{array}\right)\, ,
\end{equation}
hence dividing the number of variational parameters by $2$ for similar accuracy.

Of course this works only if the ground state does not spontaneously break $\mathbb{Z}_2$ symmetry. This is not expected to happen, and I verified at moderate bond dimensions by carrying simulations without explicit $\mathbb{Z}_2$ symmetry enforced, and noticing it was approximately restored by the optimization.

\section{Energy density}

\subsection{Analytic results}
The energy density (in a particular renormalization scheme) is not \emph{a priori} the most physically interesting observable.  However, for a variational method, it gives a very good proxy for the quality of the approximation.

For the SG and ShG models, an exact value was found by Lukyanov and Zamolodchikov \cite{lukyanov1997} and is recalled in KLM \cite{konik2021}. It reads
\begin{equation} \label{eq:e0_exact}
    \varepsilon^\text{rq}_0  = \frac{\pi M_\text{ShG}^2}{2 \sin\left[\pi(b +1/b)\right])} \, ,
\end{equation}
where $M_{\text{ShG}}$ is the Sinh-Gordon mass gap which admits the exact expression
\begin{equation}
   M_\text{ShG}(b) = \frac{4\sqrt{\pi}}{\Gamma\left(\frac{1}{2+2b^2}\right)\Gamma\left(1+\frac{b^2}{2+2b^2}\right)} \left[-\mu_\text{ShG} \pi \frac{\Gamma(1+b^2)}{\Gamma(-b^2)}\right]^\frac{1}{2+2b^2} \, ,
    \label{eq:mass}
\end{equation}
where recall that $\mu_\text{ShG} = \frac{m^{2+2b^2}}{2^{4+2b^2}\pi b^2} \e^{2b^2 \gamma_\text{E}}$. 

One gets the same expressions for the mass gap $M_\text{SG}$ of the Sine-Gordon model by replacing $\mu_\text{ShG}$ by $\mu_\text{SG}$ [given in \eqref{eq:musg}] and replacing $b$ by $ib$ in the mass formula \eqref{eq:mass}. In fact, the result was historically obtained first for the Sine-Gordon model, and formula \eqref{eq:mass} is merely an analytic continuation. This is important to have in mind, because for the Sine-Gordon model, $M_\text{SG}$ corresponds to the mass of the first breather, which seizes to exist when $b\rightarrow 1/\sqrt{2}$ as the model undergoes a phase transition from attractive to repulsive phase. At this point, the energy density in the SG ground state collapses to $-\infty$. Hence, for the SG model, the energy density for $b\in [1/\sqrt{2},1[$ given by formula \eqref{eq:e0_exact} is not that of the Hamiltonian $H_{SG}$, which is strictly speaking not bounded below. Rather, it corresponds to $H_{SG}$ minus an infinite multiple of the identity, a renormalized Hamiltonian which cannot be trivially written in the free Fock space without cutoff.

For the Sinh-Gordon model, the energy density given by formula \eqref{eq:e0_exact} remains lower bounded for $b\in[0,1]$ and goes to $0$ when $b\rightarrow 0$. Formula \eqref{eq:e0_exact} clearly seizes to make sense for $b\geq 1$ (it becomes complex). However, even if the formula \emph{a priori} makes sense for all $b\in[0,1]$ the analytic continuation could in principle break down before, for example at $b = 1/\sqrt{2}$.

\subsection{The Sine-Gordon model}
\begin{figure}
    \centering
    \includegraphics[width=0.49\textwidth]{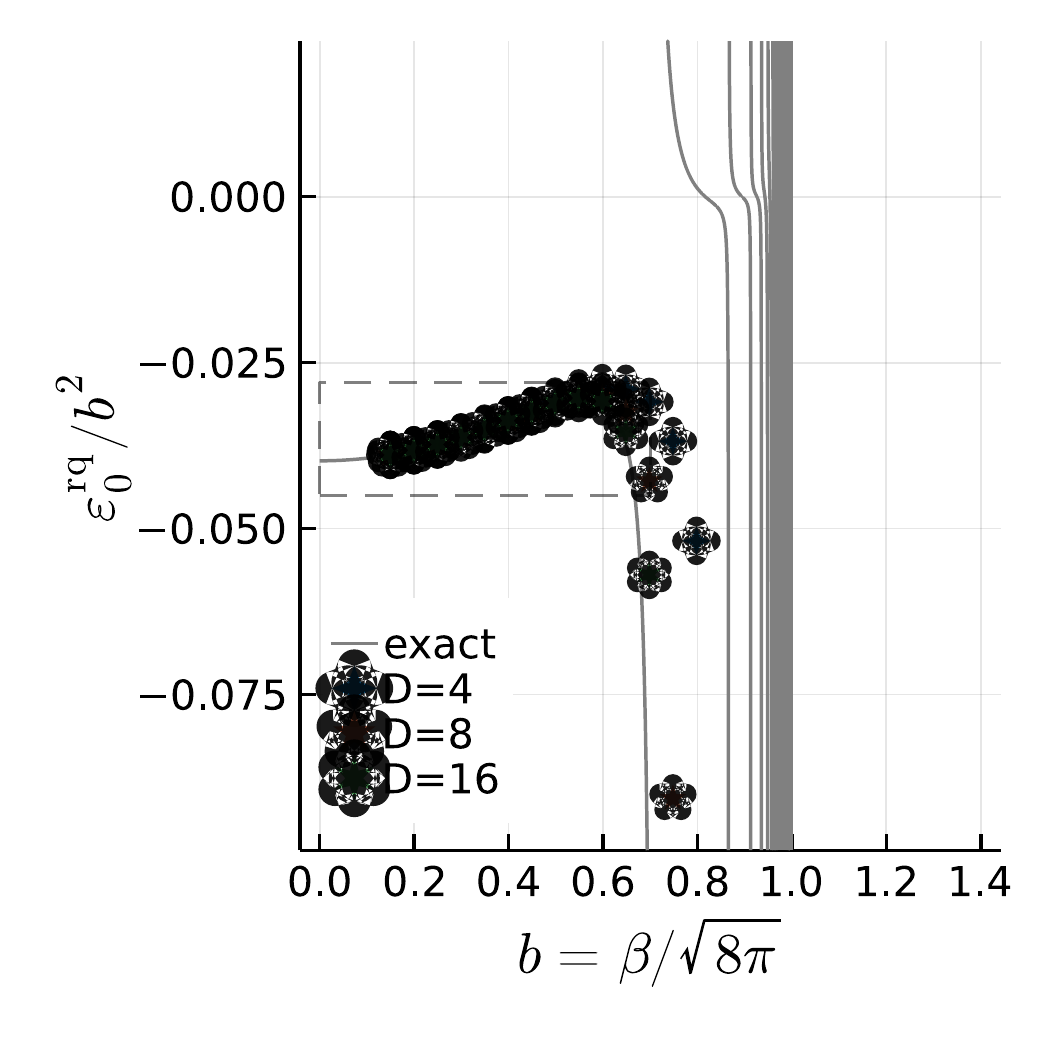}
    \includegraphics[width=0.49\textwidth]{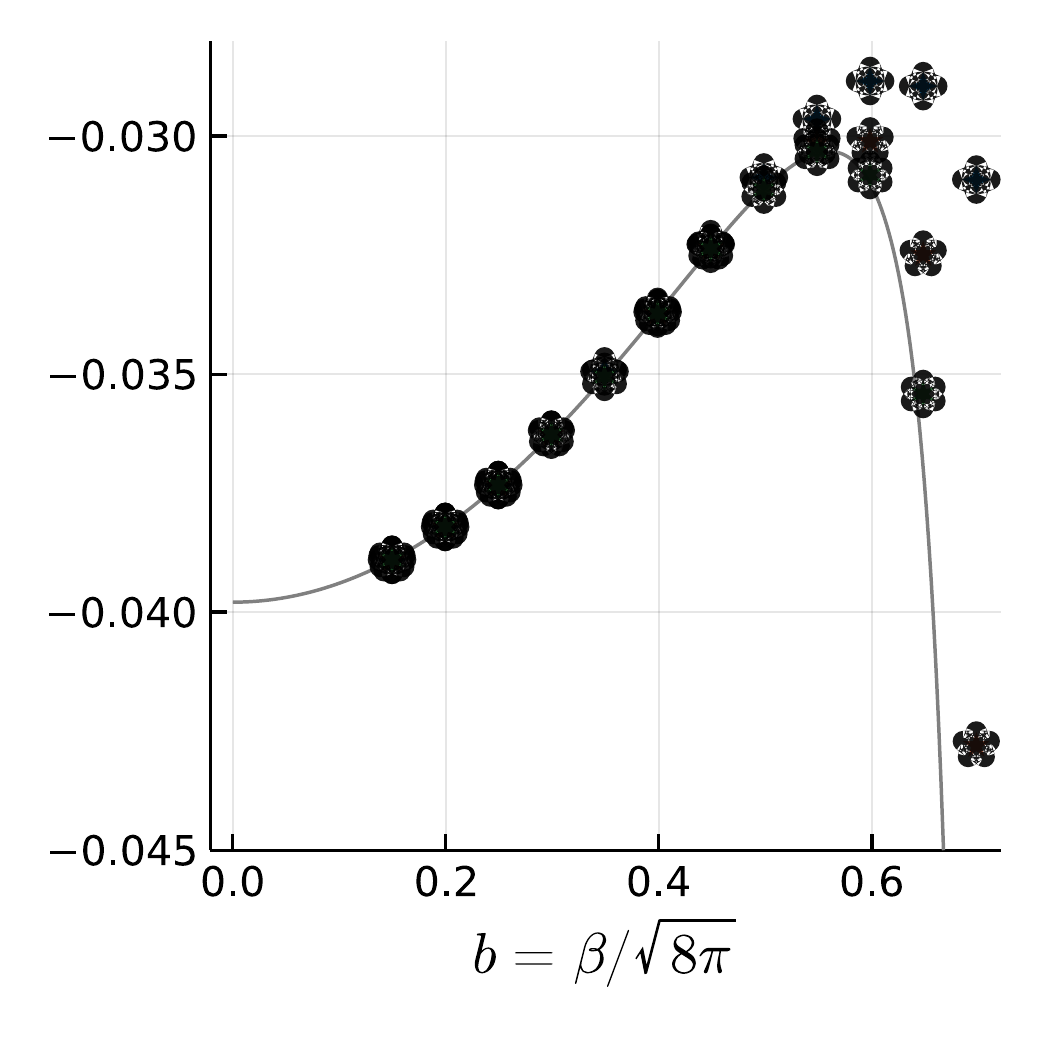}
    \caption{Rescaled ground state energy density $\varepsilon_0^\text{rq}/b^2$ of the \textbf{Sine-Gordon} model (in radial quantization). The dashed box on the left, corresponding to the region $b\in[0,1/\sqrt{2}[$ of RCMPS approximability, is magnified on the right. Note that the exact values of the energy for $b\in]\sqrt{2},1[$ do not correspond to $h_\text{SG}$, but only to the resulting finite part after additional renormalization.}
    \label{fig:energy_density_sg}
\end{figure}

To find the ground state of the Sine-Gordon model, one simply minimizes the Hamiltonian density $h_{SG}$  \eqref{eq:HSG} in equal-time quantization on a RCMPS manifold of fixed bond dimension. This gives a state $\ket{Q,R}$ and its associated energy density $\bra{Q,R} h_\text{SG} \ket{Q,R} \gtrsim \varepsilon_0^\text{etq}$. Then, to compare with the exact formula \eqref{eq:e0_exact}, one can follow KLM and use \eqref{eq:energy_dictionary} to translate the numerical (equal time) RCMPS results to radial quantization. Results are shown in Fig. \ref{fig:energy_density_sg}.

Even for fairly low dimensional manifolds (small $D$), RCMPS give an excellent approximation of the ground state, at least for values of $b$ sufficiently away from the phase transition point $b_c = 1/\sqrt{2}$. For $b = 1/\sqrt{2} + \epsilon$, small values of $D$ still give a stable optimization and a very low but finite energy density (even though the Hamiltonian is no longer lower bounded). However, and as expected, I observed some runaway optimization for larger values of $D$. In this unstable case, the minimization gives lower and lower energies at every step, with a gradient that never converges to zero. 

Note that Coleman's variational argument \cite{coleman1975}, which uses free massive ground states as ansatz, shows that the Hamiltonian density is not lower bounded for $b > 1$. Although the present approach is purely numerical, we see that it can extend Coleman's variational argument to lower values of the coupling $b\in[1/\sqrt{2},1]$. Namely, we observe numerically that a RCMPS ansatz can be tuned to get arbitrarily low values of the energy density $\langle h_\text{SG} \rangle_{R,Q}$ for $b\in[1/\sqrt{2},1]$. This is the same energy density Coleman considered, normal-ordered but not further renormalized.

Because of the lack of lower bound for $h_{SG}$, RCMPS cannot be used to directly study the $b \in [1/\sqrt{2},1]$ phase. One would need a modification of the ansatz incorporating precisely the right additional UV divergence, such that energy expectation values are finite after the infinite subtraction. Note that this is precisely what happens for Gaussian continuous tensor networks approximating interacting Bose gases in $d=2+1$ dimensions \cite{karanikolaou2020gaussian}. However, this is not realizable with the present ``plain'' RCMPS ansatz, as all its normal-ordered expectation values are finite \emph{before} the infinite subtraction.

To summarize, we get an excellent approximation of the Sine-Gordon ground state with RCMPS for coupling constants $b\in[0,1/\sqrt{2}[$. For larger values, minimization over RCMPS fails but for a fairly obvious reason: without further infinite renormalization, the normal-ordered Hamiltonian is not lower bounded.

\subsection{The Sinh-Gordon model}

\begin{figure}
    \centering
    \includegraphics[width=0.57\textwidth]{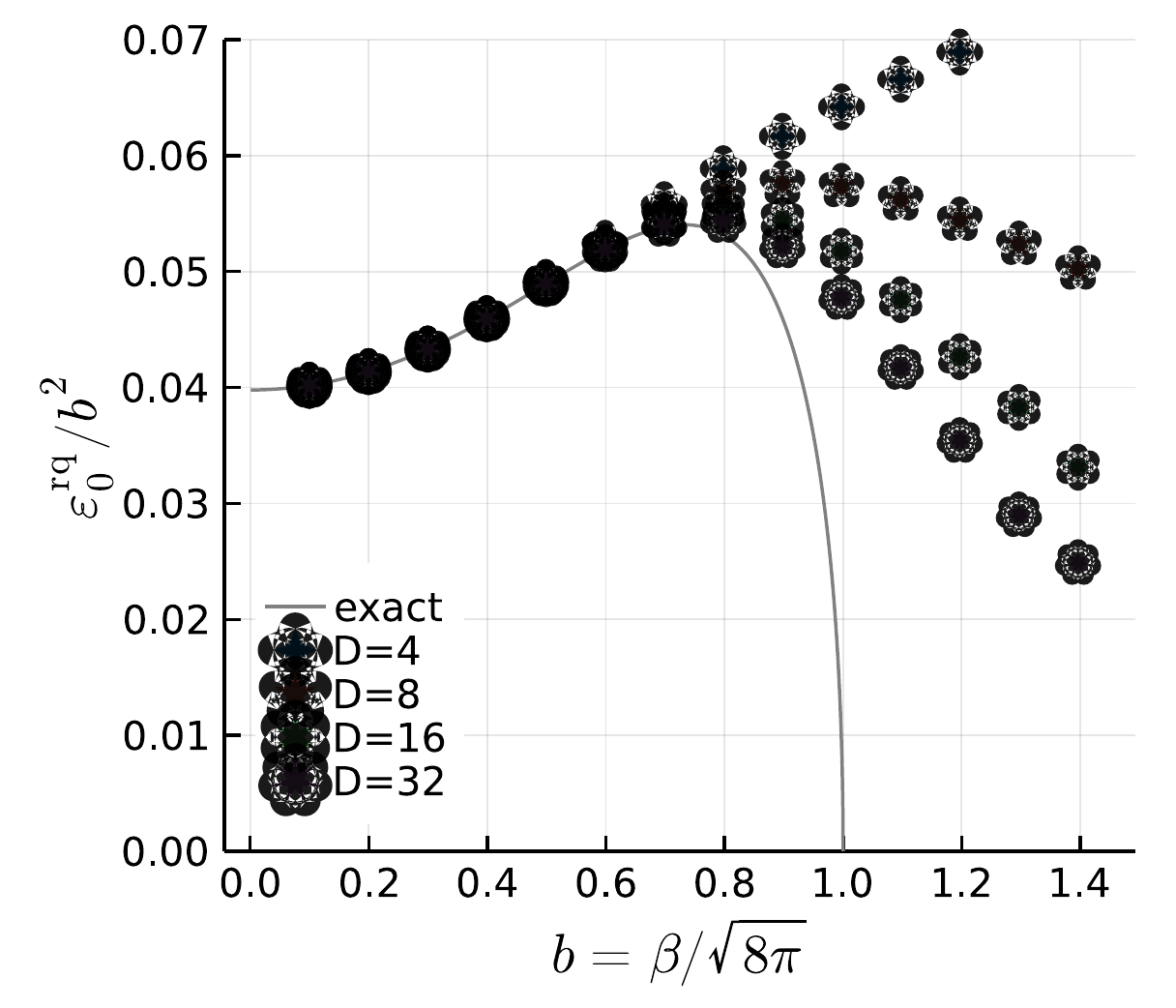}
    \includegraphics[width=0.42\textwidth]{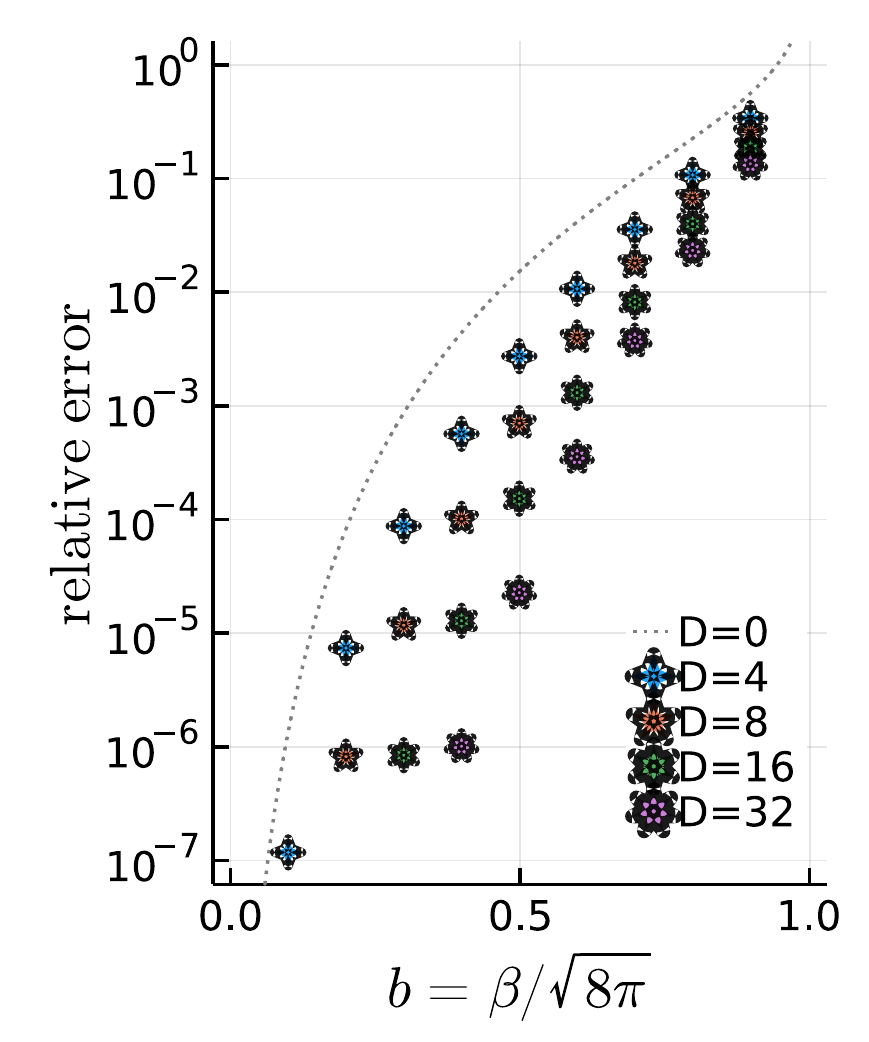}
    \caption{Rescaled ground state energy density $\varepsilon_0^\text{rq}/b^2$ of the \textbf{Sinh-Gordon} model (in radial quantization).}
    \label{fig:energy_density_shg}
\end{figure}

For the Sinh-Gordon model, the energy minimization over a RCMPS manifold with fixed $D$ is well behaved for all values of the coupling, \ie\, even for $b\geq 1$, because the energy density $\varepsilon_0^\text{rq}$ is always lower bounded by zero. The results are shown in Fig. \ref{fig:energy_density_shg} for bond dimensions up to $D=32$ and coupling constants $b\in[0.1,1.4]$.

As with the Sine-Gordon model, the approximation is good, visually sticks to the exact value, and improves quickly with $D$ at least until $b\approx 1/\sqrt{2}$. Then not only does the RCMPS upper bound substantially deviates from the exact value, but improvement as $D$ increases is slower. At least visually, on the energy density plot, one may suspect a transition of approximability around $1/\sqrt{2}$. This would also be consistent with what I observed in the optimization time: the Riemannian LBFGS minimization algorithm becomes drastically slower, requiring many more iterations, for $b \gtrsim 1/\sqrt{2}$. Plotting the relative error instead hints at a crossover, with an exponential loss of precision as $b$ increases for fixed $D$.

For $b\geq 1$, the reasonable guess is that the exact value is $\varepsilon^{\text{rq}}_0 = 0$. Indeed, if formula \eqref{eq:e0_exact} is correct, $\varepsilon^{\text{rq}}_0$ reaches $0$ at $b=1$, but $\varepsilon^{\text{rq}}_0$ is also lower-bounded by $0$, and may just saturate at $0$. The numerical RCMPS results, although far from converged in $D$, are at least consistent with this hypothesis. They also provide a rigorous energy density upper bound in this controversial region.

\section{Vertex operators expectation values}

\subsection{One-point functions}
The simplest observable one can look at is the expectation value of a vertex operator $G(a)=\langle:\!\e^{a\varphi}\!:\rangle$
It is given exactly (in radial quantization) by the Fateev-Lukyanov-Zamolodchikov-Zamolodchikov (FLZZ) formula \cite{fateev1998,lukyanov1997}, that can be analytically continued following KLM \cite{konik2021} to give, for the Sinh-Gordon model:
\begin{equation}
    \begin{split}
        G(a) = M_\text{ShG}^{-2a^2} \left[\frac{\Gamma(\frac{1}{2+2b^2}) \Gamma(\frac{2+3b^2}{2+2b^2})}{4\sqrt{\pi}}\right] \int_{\mathbb{R}^+} \!\!\frac{\upd t}{t} \left[2 a^2 \e^{-2t} - \frac{\sinh^2(2abt)}{2\sinh(b^2t)\sinh(t)\cosh(t+b^2t)}\right]\, .
    \end{split}\label{eq:flzz}
\end{equation}
This formula makes sense only up to the Seiberg bound $a \leq \frac{1}{2}\left(b + b^{-1}\right)$ after which the integral is no longer convergent. One could think of analytically continuing it to larger values of $a$, but as KLM showed, one gets negative values that do not make sense since $\langle:\!\e^{a\varphi}\!:\rangle \geq 0$ for $a$ real.

\begin{figure}
    \centering
    \includegraphics[width=0.378\textwidth]{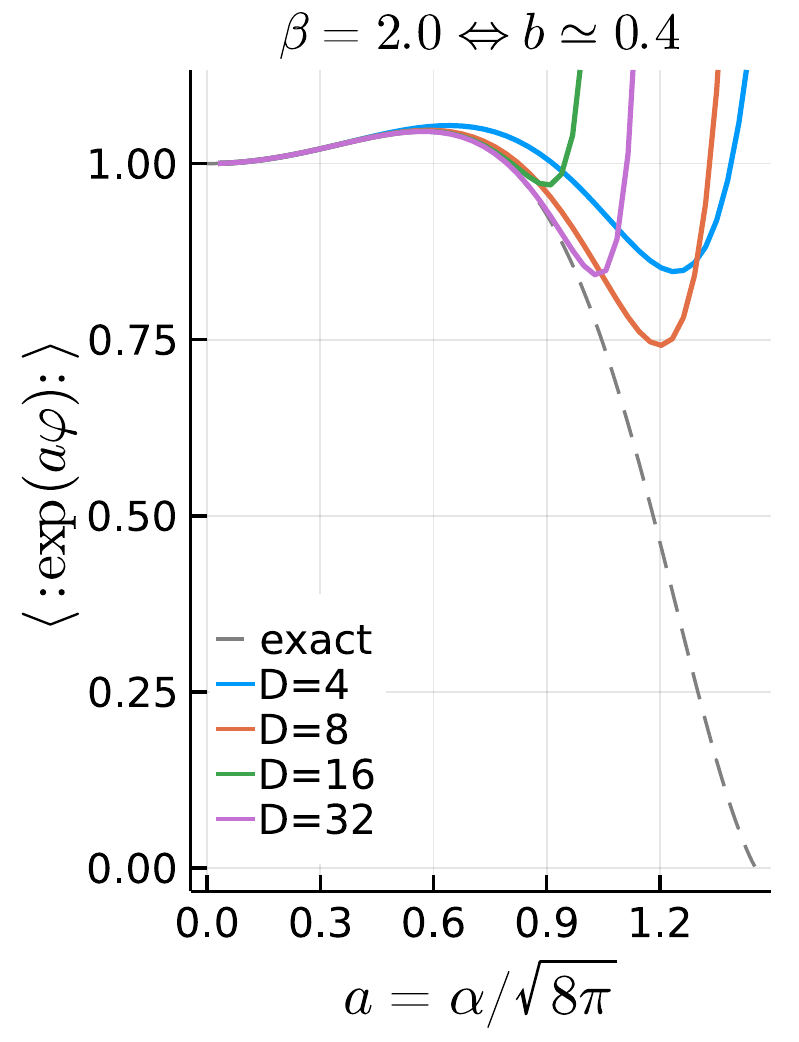}
    \includegraphics[width=0.304\textwidth]{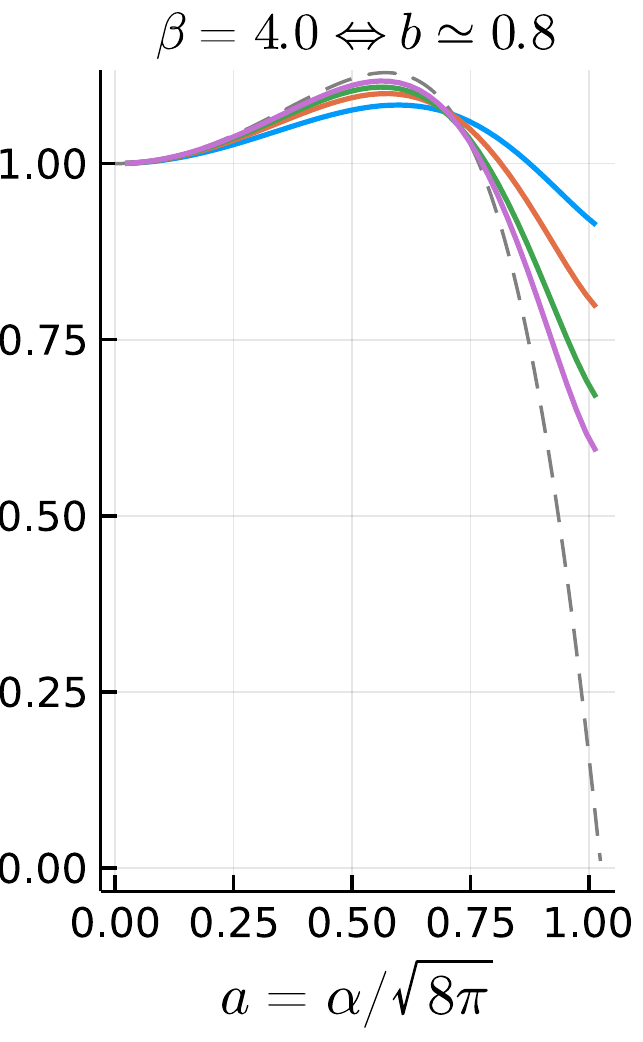}
    \includegraphics[width=0.304\textwidth]{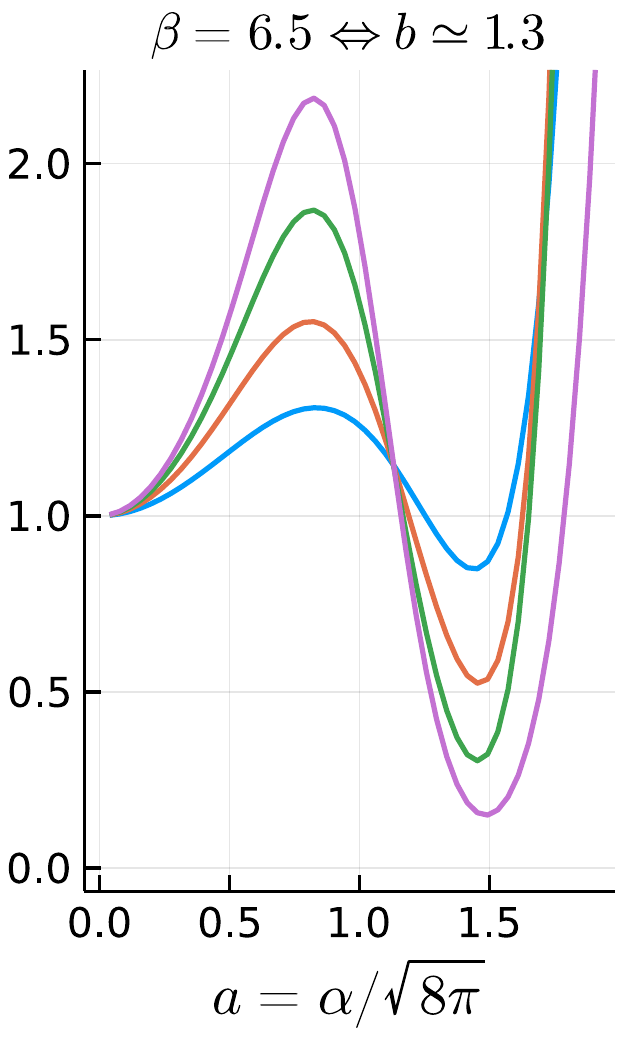}
    \caption{Vertex operator expectation values $G(a) = \langle:\!\e^{a\varphi}\!:\rangle$ for coupling constants $b=0.4, 0.8$, and $1.3$. }
    \label{fig:one_point}
\end{figure}

With RCMPS, we have access to an approximation of the ground state, and once it is known, evaluating expectation values of local operators in equal time quantization is cheap. We thus get easy access to $\langle :\! \e^{\alpha \phi(0)} \! :_m\rangle$ which we can then relate to $G(a)$ using \eqref{eq:vertex_equivalence}. The results for $3$ different values of the coupling ($b\simeq 0.4,0.8,1.3$) are given in Fig. \ref{fig:one_point} and correspond to the 3 main behaviors observed.

For values of the coupling in the ``safe'' region $b=0.4\in [0,1/\sqrt{2}]$, convergence as a function of the bond dimension is extremely fast for operators $:\e^{a\varphi}\! :$ with $a$ not too large (approximately $a\leq 2 b$). For larger $a$, still within the Seiberg bound, the behavior becomes extremely erratic: vertex operators do not get closer to the prediction of the FLZZ formula, and even seem to diverge. This behavior is reproducible, and occurs for independent successful energy minimizations terminated with vanishing gradient. Note that for $D=32$, the RCMPS we use to compute the vertex operators is intuitively close to the true ground state, differing in energy density only by a $10^{-6}$ relative error. It is thus possible that minimizing the energy density is insufficient to give finite values to large $a$ vertex operators in the thermodynamic limit. A possibility is that the FLZZ formula simply has a smaller domain of validity than previously believed, at least when the ShG model is defined via its Hamiltonian \eqref{eq:HShG}. Another option is that by taking the thermodynamic limit first and the large $D$ limit after, we fail to control some vacuum expectation values involving soft modes having vanishing contribution to the energy.

For values in the strong coupling region ($b=0.8$ in Fig. \ref{fig:one_point}), closer to the self dual point, convergence is slower (as expected) as a function of the bond dimension $D$, but seems more uniform in $a$, possibly even in the whole ``Seiberg allowed'' region. This is likely because minimizing the energy density controls vertex operators with $a$ of the order of the coupling $b$.

Finally, for ultra strong coupling, \ie\, for values beyond the self-dual point ($b=1.3$ in Fig. \ref{fig:one_point}), we enter \textit{terra incognita}. There is no reference analytical value (even speculative) to compare our numerics to, and the very existence of the model defined by its Hamiltonian $H_\text{ShG}$ is unclear. For the bond dimensions I could probe numerically, it is hard to know if the vertex operators converge very slowly to some large value, or if they do not converge at all.

\subsection{Two-point correlation functions}

\begin{figure}
    \centering
    \includegraphics[width=0.378\textwidth]{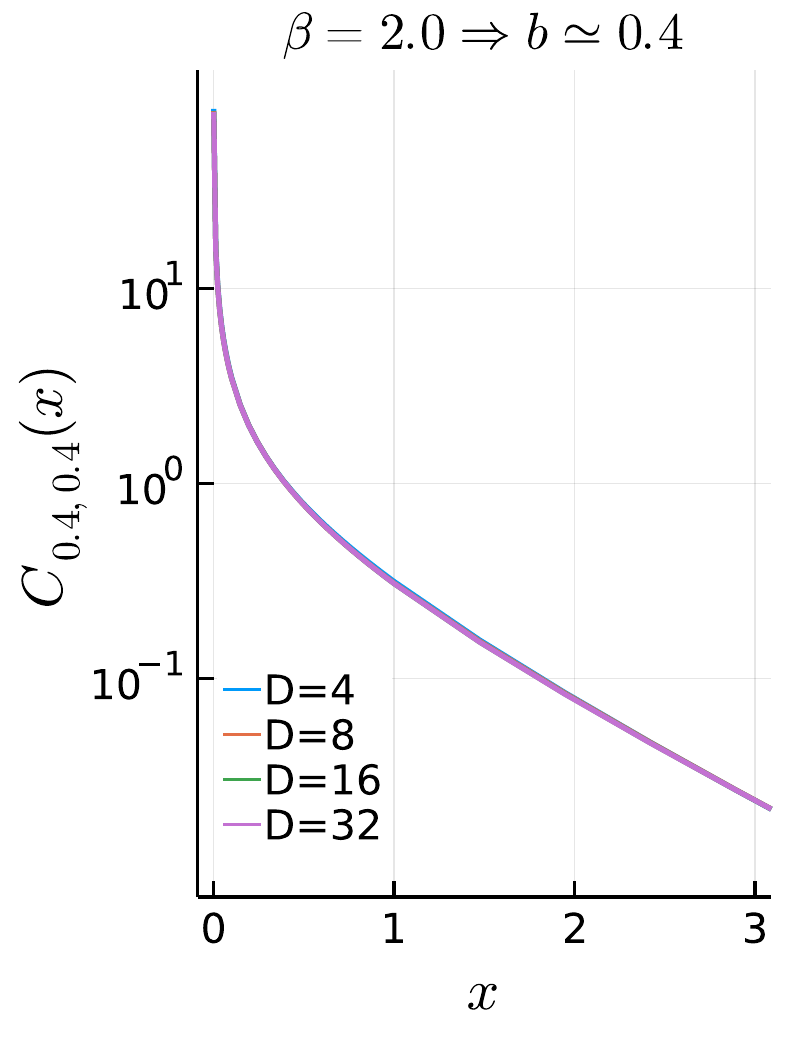}
    \includegraphics[width=0.304\textwidth]{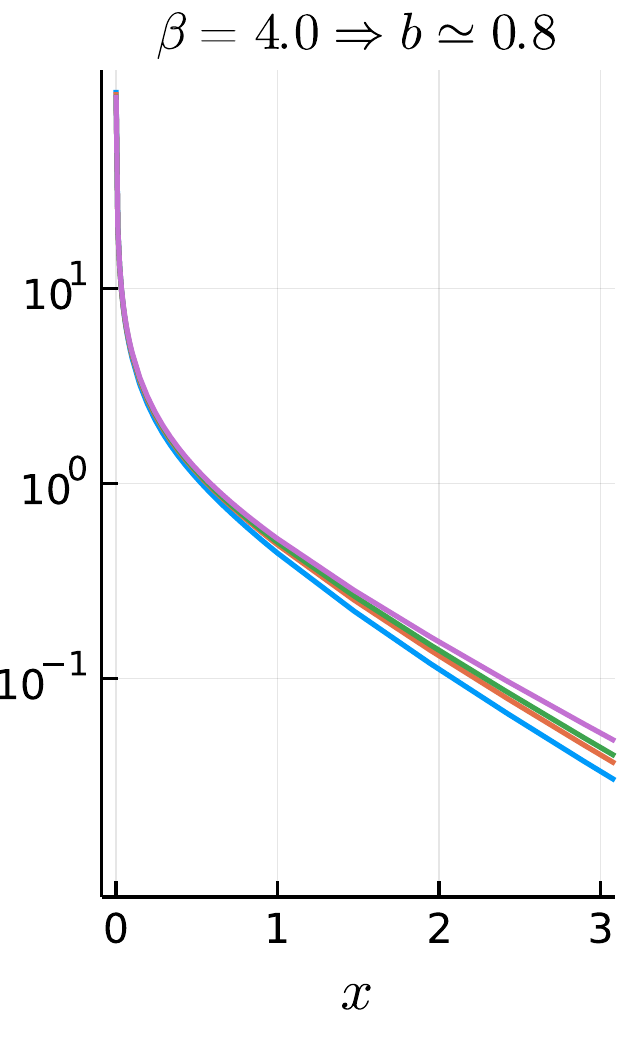}
    \includegraphics[width=0.304\textwidth]{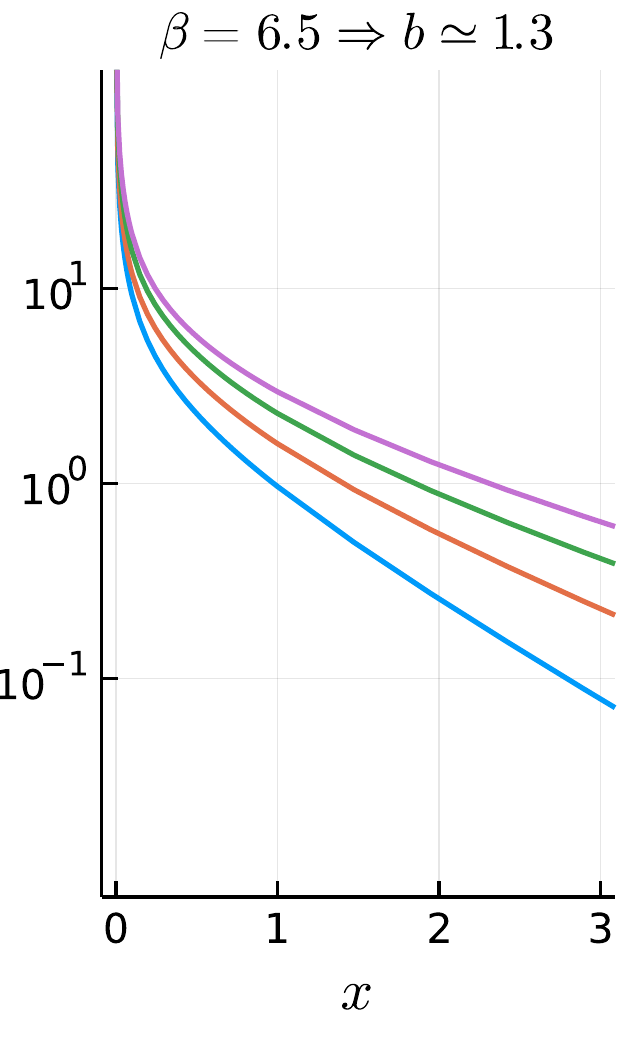}
    \caption{Vertex operator expectation values $G(a) = \langle:\!\e^{a\varphi}\!:\rangle$ for coupling constants $b=0.4, 0.8$, and $1.3$. }
    \label{fig:two_points}
\end{figure}

Once a RCMPS is given, one can easily compute all $N$-point functions of vertex operators (or field monomials) at equal-time (but not necessarily equal positions) at a cost only $\propto D^3$. These can then be converted to radial quantization conventions. As an illustration, we can consider connected two-point functions of the form:
\begin{equation}
    C_{a_1,a_2} (x) = \langle :\e^{a_1\varphi(x)}\!: \; :\e^{a_2\varphi(0)}\! :\rangle - \langle :\e^{a_1\varphi(x)}\!: \rangle\langle :\e^{a_2\varphi(0)}\! :\rangle \, .
\end{equation}
These correlation functions are not computable exactly from integrability with current techniques, and we consequently do not have a point of comparison. But it is still possible to verify that the functions converge as $D$ is increased. Illustrative results for $a_1=a_2=0.4$ are shown in Fig. \ref{fig:two_points}, but taking other values of $a_1,a_2$ gave qualitatively similar results.

As expected, convergence as a function of $D$ is slower as the coupling $b$ is increased. In all cases, the power law divergence at short distance is well captured. At large distance, the decay is exponential for $b \in [0,1[$. For $b\geq 1$, the numerical results are largely inconclusive, as the large distance behavior is not converged in $D$. However, the consistent reduction of the slope in log-scale as $D$ is increased is consistent with a massless phase as conjectured by KLM \cite{konik2021} and BLC \cite{bernard2022} (but would also be consistent with an ill-defined model).

\section{Discussion}

\subsection{Entanglement entropy and expressiveness of RCMPS}

The convergence of the observables as a function of $D$ we obtained for the Sinh-Gordon model gets dramatically worse as the coupling constant is increased. This is to be contrasted with what happened \eg\, with the $\phi^4$ model, where convergence was fast for all values of the coupling (even $\gg 1$, deep in the symmetry broken phase), at the exception of a narrow band around the symmetry breaking point. This behavior mimics what was observed by KLM with Hamiltonian truncation techniques. It likely means that the free Fock basis becomes inefficient for approximation when the coupling constant is large, since both methods, RCMPS and HT, rely on it as a starting point. 

A good way to understand this phenomenon, at least in the RCMPS context, is to look at entanglement. Standard entanglement entropy is UV divergent for relativistic field theories, which is actually one of the reasons why standard tensor network techniques become inaccurate as the UV cutoff is lifted. Here, we can consider a more exotic definition of entanglement entropy, with a different notion of locality. It is \emph{finite} and adapted to the study of the expressiveness of RCMPS. 

In the standard definition of spatial entanglement, one cuts the real line into two subsets \eg\, $x\leq$ and $x > 0$. One then splits the total Hilbert space into two factors $\mathscr{H} = \mathscr{H}_{x\leq 0}\otimes \mathscr{H}_{x>0}$, ``naturally'' associated to each side. What is natural in the relativistic context is to pick the Fock spaces associated to $\hat{\phi}(x),\hat{\pi}(x)$, that is informally the space of wavefunctions $\psi(\phi)$ for fields $\phi$ with supports localized on either side. One then gets entanglement entropy of a state on the half-line by considering the Von Neumann entropy of the reduced density matrix 
\begin{equation}
\rho_{x> 0} = \tr_{\mathscr{H}_{x\leq 0}} \left( \ket{\psi}\bra{\psi}\right)\, .
\end{equation}
This definition is the right one if one demands that the notion of spatial locality used to define entanglement match the locality of the dynamics given by the Hamiltonian. The main drawback of this definition is that, as is well known, it gives a  logarithmically UV divergent entanglement entropy, which can be understood from the fact that a relativistic theory (even a massive one) is a CFT at short distances. 

\begin{figure}
    \centering
    \includegraphics[width=0.49\textwidth]{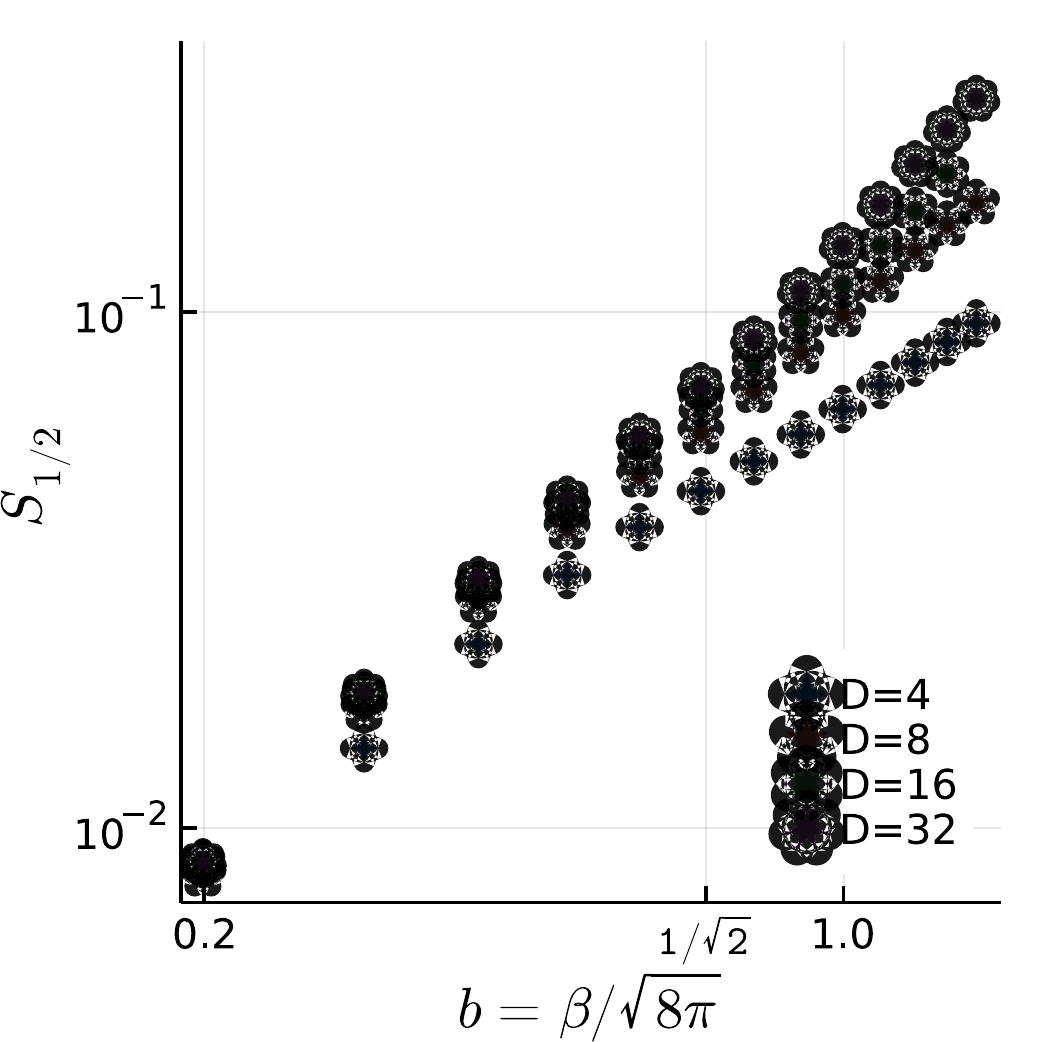}
    \includegraphics[width=0.49\textwidth]{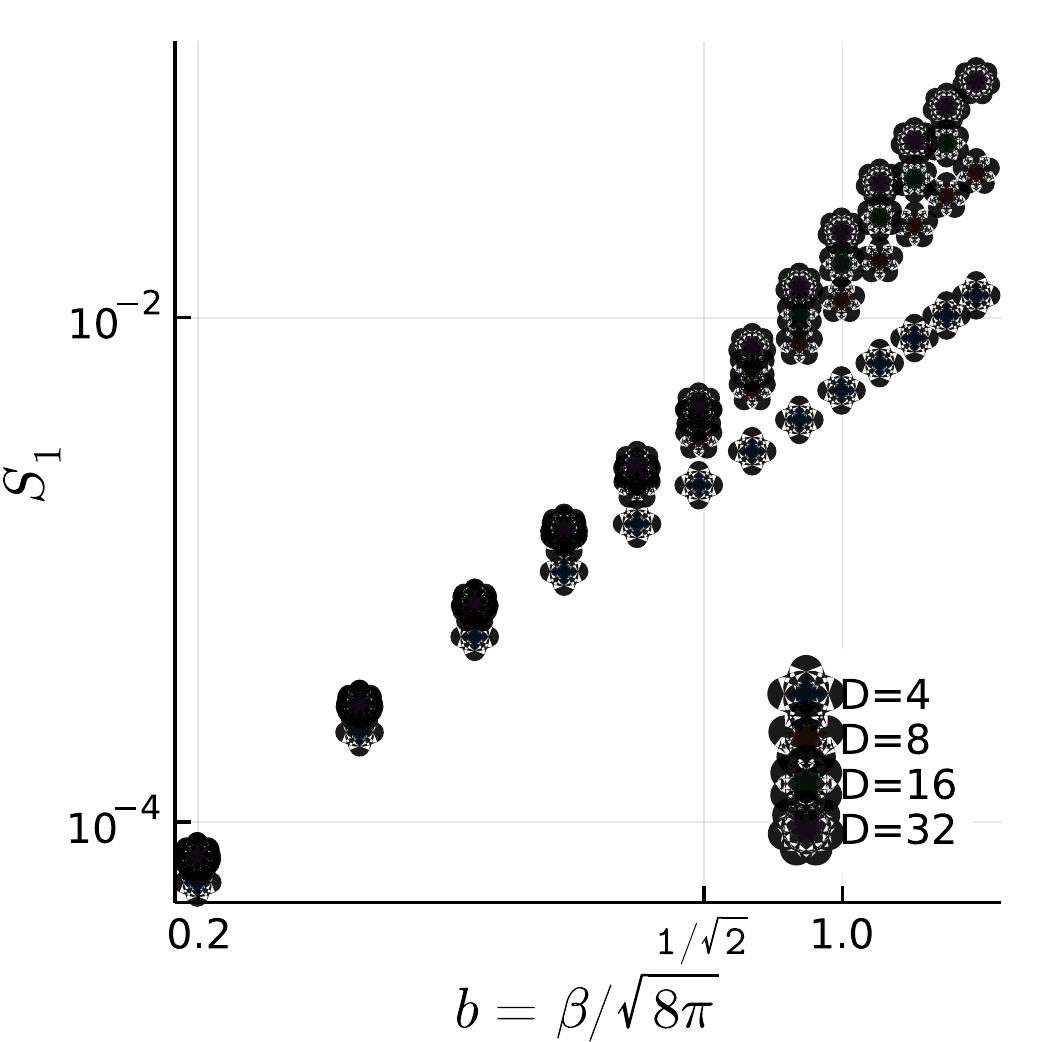}
    \caption{Half-line entanglement in the free basis for the ground state of the Sinh-Gordon model. Left: $1/2$ Renyi entropy $S_{1/2}$. Right: Von Neumann entropy $S_1$. }
    \label{fig:entanglement}
\end{figure}

However, here we care primarily about entanglement as a measure of approximability with tensor network states. The natural choice is thus to consider a new notion of spatial locality which is the one associated to our tensor decomposition, and not the one implied by the locality structure of the Hamiltonian. Namely, we consider the tensor product decomposition into Fock spaces associated to $a(x), a^\dagger(x)$. Since the field $\phi(x)$ is obtained from $a,a^\dagger$ via a convolution (a non-local change of basis), this is indeed a different notion of locality. As is well known with matrix product states, the whole entanglement spectrum for such bipartitions is contained in the stationary state of the transfer operator. For RCMPS, this state is simply $\rho_0$ where $\rho_0$ is such that $\mathcal{L}\cdot \rho_0 = Q\rho_0 + \rho_0 Q^\dagger + R\rho_0 R^\dagger = 0$ (in left canonical gauge). With this new definition, entanglement entropy in the free ground state is exactly zero. As this divergent contribution is removed, the remaining entanglement for an interacting theory appears to be finite (see in Fig. \ref{fig:entanglement}) at least for small enough coupling.

In the case of the ShG model, we observe that this new entanglement entropy increases polynomially in the coupling $b$. For $b\geq 1/\sqrt{2}$, our estimate is manifestly not converged. Either the entanglement entropy remains finite but becomes so large that much higher bond dimensions would be needed to estimate it, or there is a phase transition to infinite entanglement for some threshold coupling $b_c$ (that could plausibly be $b_c=1/\sqrt{2}$). In any case, whether it is simply a fast growth or a divergence, this increase of the entanglement entropy in the free particle basis explains why both HT and RCMPS struggle to capture accurately the properties of the ground state as the coupling is increased.

\subsection{Possible improvements}
The present study could be improved in many ways, which would likely help give a clearer picture of the ShG model. 

The first thing that comes to mind is simply to improve the precision of the numerical results. This could be done by using larger bond dimensions. This should be possible by improving the routines used to evaluate expectation values and their gradient: using ODE solvers adapted to non-autonomous linear problems or using a GPU to evaluate the generator. One can also likely reduce the number of evaluations required for the energy minimization: using better initializations (from lower $D$ results instead of random), building a quasi-Newton algorithm adapted to the specific form of our problem, or finding a way around the inversion of $\rho_0$ to avoid regulators that harm performance. Finally, one could improve precision by doing extrapolations in $D$, similarly to what is done in HT with extrapolations in the truncation energy $E_T$ \cite{konik2021}. In the present paper, I refrained from doing so to keep the variational nature of the results, but this would dramatically improve precision.

Computing a wider range of observables would also be desirable. For example, one should be able to directly extract the mass gap by minimizing the energy on the tangent space of the RCMPS approximate ground state \cite{vanderstraeten2019tangentspace}. One should also be able to compute approximate $c$-functions to see if the UV fixed point changes of nature at large coupling.

\subsection{Open problems}
The main open problem that is yet to be solved is to understand the exact nature of the ShG model for $b\geq 1$. The present numerical results corroborate the suspicions of KLM \cite{konik2021} and BLC \cite{bernard2022}, that the model is massless for $b\geq 1$, but are clearly insufficient to conclude.

To make progress, I believe the first step is to understand if the model we are working with, defined from its normal-ordered Hamiltonian, is well defined (or can be made well defined) even only for $1/\sqrt{2} \leq b \leq 1$ (and then of course settle the $b\geq 1$ case). If the model turns out to be sick for large values of the coupling, is there an additional renormalization step one can do to cure it? The recent rigorous probabilistic constructions of Liouville theory with Gaussian multiplicative chaos \cite{david2016,vargas2017} give hope that such a clean understanding could be obtained. Then, one could probe the physics more confidently, without worrying that the peculiar behavior we see is just an artifact of some ill-definiteness. 

On the tensor network side, an open problem is to understand how to deal with divergences that remain after normal-ordering as in the Sine-Gordon model for $b \geq 1/\sqrt{2}$ (and perhaps also for the Sinh-Gordon model). Ideally, one would want to do that without introducing a new cutoff parameter, which would thus require modifying the RCMPS ansatz we have used (since the latter necessarily gives finite expectation values for normal-ordered Hamiltonians). 

\subsection*{Acknowledgments} I am grateful to Jheng-Wei Li, M\'arton L\'ajer, Robert Konik,  G\'abor Tak\'acs, and Denis Bernard for helpful discussions about the Sinh-Gordon model. I would also like to thank Giuseppe Mussardo whose presentation 4 years ago at IHES sparked my curiosity about this model. Finally, I am indebted to Jutho Haegeman and other members of the quantum group at the university of Ghent: I and made an extensive use of  \texttt{OptimKit.jl} and \texttt{KrylovKit.jl}, two wonderful Julia packages they developed. 

\vskip2cm
\appendix 
 \section{Computing RCMPS expectation values}\label{app:expectations}

 The computation of RCMPS expectation values relies on standard CMPS techniques (see \eg\, \cite{haegeman2013}) and on recently introduced tricks that are specific to RCMPS (see \cite{rcmps_article}). To make the present article self-contained, all the steps required are explained in detail here.

\subsection{Computing the norm}
For a RCMPS 
\begin{equation}
    \ket{Q,R} = \tr\left\{\mathcal{P} \exp\left[\int_0^L \upd x \, Q\otimes \mathds{1} 
    + R\otimes a^\dagger(x)\right]\right\}\ket{0}\, ,
\end{equation}
the very first observable to compute is the identity, \ie\, the norm $\bra{Q,R} Q,R\rangle$. The result is
\begin{equation}
    \bra{Q,R}Q,R\rangle=\tr_{\mathbb{C}^D\otimes \mathbb{C}^D}\left[\exp (L \mathbb{T})\right]
\end{equation}
with the transfer operator $\mathbb{T}$
\begin{equation}
    \mathbb{T} = Q\otimes \mathds{1} + \mathds{1} \otimes Q^* + R\otimes R^* \,.
\end{equation}
There are many ways to prove this continuum result, which is standard (see \eg\, ). To make this appendix self-contained, we can recall the simplest, although not the most elegant way, which is to discretize space\footnote{Note that discretization is not needed \emph{a priori}, and one can work with RCMPS without ever discretizing space}. This gives, for a lattice spacing $\varepsilon$
\begin{equation}\label{eq:cmps_bj}
    \ket{Q,R}=\tr\left\{\prod_{j=1}^{L/\varepsilon} \exp\left[\varepsilon Q + \sqrt{\varepsilon} R\, b^\dagger_j \right] \right\}\ket{\Omega} + O(\varepsilon) \,
\end{equation}
with $b_j:= \frac{1}{\sqrt{\varepsilon}} \int_{j\varepsilon}^{(j+1)\varepsilon} a(x)$,
which is indeed a good discrete bosonic operator with $[b_j,b_k^\dagger]=\delta_{ij}$.
The norm then reads
\begin{align}
    \bra{Q,R} Q,R\rangle &=\bra{\Omega} \tr \left\{\prod_{j=1}^{L/\varepsilon} \exp\left[\varepsilon Q^* + \sqrt{\varepsilon} R^*\, b_j \right]\right\}\tr\left\{ \prod_{j=1}^{L/\varepsilon}\exp\left[\varepsilon Q+ \sqrt{\varepsilon} R\, b^\dagger_j \right] \right\}\ket{0} \\
    &=  \tr\left\{\prod_{j=1}^{L/\varepsilon} \bra{0_j}\exp\left[\varepsilon \mathds{1}\otimes Q^* + \sqrt{\varepsilon} \mathds{1}\otimes R^*\, b_j \right]\exp\left[\varepsilon Q\otimes \mathds{1} + \sqrt{\varepsilon} R\otimes \mathds{1}\, b^\dagger_j \right] \ket{0_j} \right\}
\end{align}
Here, we put everything into the same trace using that $\tr[A\otimes B] = \tr[A]\tr[B]$. Further, we used that the bosonic vacuum $\ket{0}$ is simply a tensor product of states $\ket{0_j}$ annihilated by the $b_j$. Finally, expanding the exponentials to order $O(\varepsilon)$ 
\begin{align}
     \bra{Q,R} Q,R\rangle &\simeq \tr\left\{\prod_{j=1}^{L/\varepsilon} \bra{0_j} 1 + \varepsilon (\mathds{1}\otimes Q^*+ Q \otimes\mathds{1} + R \otimes R^* b_j b_j^\dagger) + \sqrt{\varepsilon} (\mathds{1}\otimes R^*\, b_j +R\otimes \mathds{1}\, b^\dagger_j)   \ket{0_j} \right\}\\
     &\simeq \tr\left\{\prod_{j=1}^{L/\varepsilon} 1 + \varepsilon \mathbb{T}  \right\}\simeq \tr\left\{\prod_{j=1}^{L/\varepsilon} \exp\left(\varepsilon \mathbb{T}\right)  \right\} \simeq \tr \left[\exp L \mathbb{T}\right]
\end{align}
In what follows, we will need extremely minor generalizations of this formula to overlaps $\bra{Q_2,R_2} Q_1,R_1\rangle$ of RCMPS with space-dependent matrices. Following the exact same steps as above gives
\begin{equation}\label{eq:state_overlap}
    \bra{Q_2,R_2} Q_1,R_1\rangle = \tr\left[\mathcal{P}\exp\left(\int_0^L \mathbb{T}_{1,2}\right)\right] \, ,
\end{equation}
with
\begin{equation}
    \mathbb{T}_{1,2}(x) = Q_1(x)\otimes \mathds{1} + \mathds{1} \otimes Q_2^*(x) + R_1(x)\otimes R_2^*(x) \, .
\end{equation}

\subsection{Thermodynamic limit, superoperator rewriting, and gauge fixing}

Before moving to the computation of other expectation values, we can now recall the standard steps, common to all matrix product states, to fix the norm to $1$ in the thermodynamic limit, reduce the cost of operations, and simplify the parameterization. 

The norm of the CMPS scales $\propto (\ell_1|r_1)\exp(\lambda_1 L)$ where $\lambda_1$, $|r_1)$, $|l_1)$ are the eivenvalue with largest real part of $\mathbb{T}$ and its associated right and left eigenvectors. We may cancel this behavior, without loss of generality, by simply substituting $Q\rightarrow Q-\lambda_1 \mathds{1}$. With this new choice, the leading eigenvalue of $\mathbb{T}$ is $0$ and all the other ones have strictly negative real part. Diagonalizing $\mathbb{T}$ one gets:
\begin{equation}
    \e^{L\mathbb{T}} = \sum_{j=1}^{D^2} \e^{\lambda_j} |r_j)(\ell_j| \underset{L\rightarrow \infty}{\longrightarrow} |r_1)(\ell_1|
\end{equation}
and thus in the thermodynamic limit, the exponential of the transfer matrix becomes a simple rank-$1$ projector.

The second step, analogous to what is done with MPS, is to go to a super-operator representation of the transfer operator. It consists in identifying the tensor-product vector space $\mathbb{C}^D\otimes \mathbb{C}^D$ with the space of complex matrices $\mathcal{M}_D(\mathbb{C})$:
\begin{equation}
    |v) = \sum_{k,l} v_{k,l} \ket{k}\otimes \ket{l} \rightarrow v = \sum_{k,l} v_{k,l} \ket{k}\bra{l} \, .
\end{equation}
With this mapping, we can introduce the super-operator $\mathcal{L}$ which reproduces the action of $\mathbb{T}$ on $v$ now written as a matrix
\begin{equation}
\mathbb{T} |v) \rightarrow \mathcal{L}\cdot v = Qv +vQ^\dagger + R v R^\dagger \,.
\end{equation}
This makes evaluation of the transfer operator scale better

The third step, like in the discrete, is to note that there is a lot of redundancy in the parameterization of the CMPS. In particular, it is straightforward to see that conjugating the matrices $Q,R$ with an invertible matrix $U$ does not change the state
\begin{equation}
    \ket{U^{-1}QU,U^{-1}RU} = \ket{Q,R}\,.
\end{equation}
This can be exploited to fix properties of $Q$ and $R$, to simplify computations without losing expressiveness. A convenient choice is the \emph{left canonical form}, which is obtained by taking $U=\ell_1$ where $(\ell_1|$ is the leading left eigenvector of $\mathbb{T}$. By definition, this matrix verifies
\begin{equation} 
\ell_1 Q +  Q^\dagger \ell_1 + R^\dagger \ell_1 R = 0
\end{equation}
Now taking $Q_\ell = C Q C^{-1}$ and $R_\ell = C R C^{-1}$ where $\ell_1=C^\dagger C$ we get
\begin{equation}\label{eq:left-canonical}
    Q_\ell + Q_\ell^\dagger + R_\ell^\dagger R_\ell = 0
\end{equation}
This implies that the identity matrix, once vectorized, is a left eigenvector of $\mathbb{T}$ with eigenvalue 0. Equivalently, this implies that $\mathcal{L}$ is of the Lindblad form. In practice, one can choose, without loss of generality, matrices verifying \eqref{eq:left-canonical} from the beginning. Writing $Q_L=-i K - R_L^\dagger R_L/2$, equation \eqref{eq:left-canonical} is equivalent to $K$ being self-adjoint. One can thus parameterize the CMPS directly with $K$ self-adjoint and $R_L$ (which we will write simply $R$).

To summarize, after these operations, we now have $\bra{Q,R}Q,R\rangle  = (\ell_1|r_1) = \tr(\rho_0) = 1$ in the thermodynamic limit, where $\rho_0$ is such that $\mathcal{L}\cdot \rho_0 = 0$. From now on we always work in this gauge.

\subsection{Correlations of $a(x)$}
The easiest operators to compute besides the norm are normal ordered expectation values of $a(x)$ and $a^\dagger(x)$, for which it is easy to construct a generating functional $Z_{j',j}$:
\begin{equation}
    \mathcal{Z}_{j',j} = \bra{Q,R}\exp\left[\int j'(x) a^\dagger(x)\right] \exp\left[\int j(x) a(x)\right] \ket{Q,R}\, .
\end{equation}
It gives all normal-ordered correlation functions of $a(x)$ operators, \eg 
\begin{equation}
    \langle a^\dagger(x)a(y)\rangle_{Q,R} = \frac{\delta}{\delta j'(x)} \frac{\delta}{\delta j(y)} \mathcal{Z}_{j',j} \bigg|_{j,j'=0} \,. 
\end{equation}
This generating functional is easily computed using the Baker-Campbell-Hausdorff (BCH) formula and the formula for state overlaps \eqref{eq:state_overlap}:
\begin{align}
    \mathcal{Z}_{j',j} &= \bra{Q,R} \exp\left[\int j(x)a(x)\right] \exp\left[\int j'(x) a^\dagger(x)\right] \ket{Q,R} \exp\left(- \int j'j\right)\\
    &=\bra{Q,R+j^*} Q,R+j'\rangle\exp\left(- \int j'j\right)\\
    &= \tr \left[\mathcal{P}\exp \left(\int \mathbb{T} +  j\,R\otimes\mathds{1} + j'\,\mathds{1}\otimes R^*\right)\right]\,.
\end{align}

\subsection{Vertex operators}
Correlations of $a(x)$ fully characterize the state, but are not directly useful in the relativistic context, where we care about expectation values of the field $\phi(x)$ itself, which are obtained from $a(x)$ through a convolution
\begin{equation}\label{eq:field_convolution}
    \phi(x) = \int \upd y J(x-y) \left[a(y) + a^\dagger(y)\right]
\end{equation}
where $J$ is a real function
\begin{align}
    J(x):&= \frac{1}{2\pi} \int \frac{\upd k}{\sqrt{2\omega_k}} \, \e^{-ikx} \label{eq:J_def}\\
    &= \frac{K_{1/4}(|x/m|)}{2^{9/4}\sqrt{\pi}\, \Gamma(5/4)\, |x/m|^{1/4}} \,
    \end{align} 
and $K_{\nu}(x)$ is the modified Bessel function of the second kind.

The second easiest operators to compute, which are useful for SG and ShG models we consider, are vertex operators
\begin{equation}\label{eq:vertex_definition}
    \langle V_\beta\rangle:=\bra{Q,R} :\! \e^{\beta\phi(x)}\!: \ket{Q,R}\,.
\end{equation}
By translation invariance, we can fix $x=0$ and using the expression of the field as a function of $a(x)$ we get
\begin{align}
    :\! \e^{\beta\phi(0)}\!: &= :\!\exp\left[ \frac{\beta}{2\pi }\int \upd x \! \int \frac{\upd k}{\sqrt{2\omega_k}} \e^{-ikx} a(x) + \e^{ikx} a^\dagger(x) \right]\!\! :\nonumber\\
    &= \exp\left[\beta \int \upd x \, J(x) a^\dagger (x) \right]\exp\left[\beta \int \upd x \, J(x) a (x)\right] \label{eq:ordered_vertex}
\end{align}
The expectation value of vertex operators is thus simply the generating functional for sources $\beta J$:
\begin{align}\label{eq:vertex_expectation}
    \langle V_{\beta}\rangle &= Z_{\beta J,\beta J}
    = \tr\left(\exp\left\{\mathcal{P}\exp\left[\int \upd x \; \mathbb{T} + \beta J(x)(R\otimes \mathds{1} + \mathds{1} \otimes R^*)\right]\right\}\right)
\end{align}
The path-ordered exponential is simply the solution of an ordinary differential equation. Moving to the superoperator representation this gives $\langle V_\beta \rangle = \tr(\rho_\infty)$ where  $\rho_\infty = \lim_{x\rightarrow \infty} \rho(x)$ and $\rho(x)$ verifies the ODE
\begin{equation} \label{eq:vertex_ode}
    \partial_x \rho(x) = \mathcal{L} \cdot \rho(x) + J(x) [R\, \rho(x) + \rho(x) \, R^\dagger]
\end{equation}
with initial condition $\rho(x = -\infty) = \rho_0$ (in fact, any positive definite matrix of trace $1$ would give the same result).

Solving this ODE takes a time proportional to the number of times the generator has to be evaluated and the cost of this evaluation (which is $D^3$). Naively, one may think that we are implicitly going back to a discretization, with an error proportional to the spacing $\varepsilon$ between points. However, for such an ODE, one can use $n$th-order Runge-Kutta schemes, with an error in $\varepsilon^n$, that reach close to machine precision with only a few thousand evaluations at order $5$ to $7$. This is one crucial advantage of working directly in the continuum, and not with a very fine lattice.

\subsection{Monomials}

When working with RCMPS, we also need expectation values of monomials $\langle :\phi^n\!:\rangle$, for example for a mass term. It is needed even for the SG and ShG models, even though they do not have an explicit mass term, because we will see that it is easier to compute expectation values of a free massive Hamiltonian rather than a massless one (and thus, we will ultimately need to subtract a mass).

To obtain such monomials, we simply differentiate vertex operators with respect to $b$
\begin{equation}
    \langle :\! \phi^n\!:\rangle= \frac{\partial^n}{\partial b^n} \langle V_\beta\rangle \bigg|_{\beta =0} \, .
\end{equation}
This allows to obtain $\langle :\phi^n\!:\rangle$ by differentiating the ODE \eqref{eq:vertex_ode}. This yields 
\begin{equation}
    \langle :\phi^n\!:\rangle = \lim_{x\rightarrow + \infty} \tr\left[\rho^{(n)}_x\right]
\end{equation}
where $\rho^{(k)}:= \partial_\beta^k \rho^b|_{\beta =0}$ obey $n$ coupled ODE
\begin{equation}
    \frac{\upd}{\upd x} \rho^{(k)}(x) = \mathcal{L}\cdot \rho^{(k)}_x + bJ(x) [R\, \rho^{(k-1)}(x) + \rho^{(k-1)}(x) \, R^\dagger]\, .
\end{equation}
Hence the cost of computing $\langle :\! \phi^n\!:\rangle$ is only $\propto n D^3$.

\subsection{Kinetic term}

The last important term to compute to obtain the energy density is the free Hamiltonian. The easiest is to compute the expectation value of the Hamiltonian density of the massive free boson $h_\text{fb}$, because it takes a particularly simple for in the $a$ basis. Indeed
\begin{equation}
    :H_\text{fb}: = :H_0\!:_m + \frac{1}{2}\int m^2\,:\phi^2\!:_m = \frac{1}{2\pi}\int \upd k \, \omega_k \; a^\dagger_k a_k\, .
\end{equation}
The corresponding Hamiltonian density $h_\text{fb}(x)$ is
\begin{align}
    :h_\mathrm{fb}(x)\!: &= \frac{1}{2\pi} \int \upd k \upd y \, \omega_k\;  \e^{ik(y-x)}a^\dagger(y) a(x) \, \\
    &= \frac{1}{2\pi} \int \upd y  \, \frac{\upd k}{\omega_k}\;  \e^{ik(y-x)} (m^2 + \partial_y \partial_x ) a^\dagger(y) a(x) \, .
\end{align}
Introducing $J(x)$ we get
\begin{equation}
    \begin{split}
    \langle :h_\mathrm{fb}: \rangle = 2 m^2 \left \langle \int \upd x  J(x) a^\dagger(x)  \!\! \int \upd y J(y) a(y)\right\rangle 
    + 2 \left\langle \int \upd x J(x) \partial_x a^\dagger(x) \!\! \int \upd y J(y) \partial_y a(y) \! \right\rangle  \, .
\end{split}
\end{equation}
This expression can easily be obtained from a derivative of two generating functionals:
\begin{align}
  \langle :h_\mathrm{fb}: \rangle  =& 2 \left[m^2 \frac{\partial}{\partial \beta_1} \frac{\partial}{\partial \beta_1} \mathcal{Z}_{\beta_1J,\beta_1 J}   + \frac{\partial}{\partial \beta_1} \frac{\partial}{\partial \beta_1} \mathcal{Y}_{\beta_1J,\beta_1 J} \right]_{b_{1,2}=0} \, ,
\end{align}
where $\mathcal{Z}$ is the generating functional for correlation functions of $a(x)$ we introduced before, and $\mathcal{Y} $ is the analog for derivatives $\partial_x a(x)$:
\begin{equation} \label{eq:def_Y}
    \mathcal{Y}_{f,g} := \bra{Q,R} \exp\left[-\int f(x) \,  \partial_x a^\dagger(x)\right] \exp\left[-\int g(x) \, \partial_x a(x)\right]\ket{Q,R}\, .
\end{equation}
It can also be evaluated exactly, and takes the same form as $\mathcal{Z}$ with the substitution $R\rightarrow [Q,R]$
\begin{equation}\label{eq:result_Y}
    \mathcal{Y}_{f,g} \! = \! \tr\left\{\mathcal{P} \exp\!
    \bigg(\int \! \, \mathbb{T}  + g\, [Q,R] \otimes \mathds{1}  +f\,\mathds{1}\otimes [Q^*, R^*]\bigg)\right\}\, .
\end{equation}
However, the proof is not completely obvious. The first step is to integrate by part in eq. \eqref{eq:def_Y} to note that $\mathcal{Y}_{f,g} = \mathcal{Z}_{\partial_x f,\partial_x g}$. One would want to integrate by part in the latter expression, to get an expression depending on $(f,g)$, but this is not trivial because of the path-ordering. Again, the inelegant way, which consists in discretizing the path-ordered exponential, provides a fairly direct proof:
\begin{align}
    \mathcal{Z}_{\partial_x f,\partial_x g} &= \! \tr\left\{\mathcal{P} \exp\!
    \bigg(\int \! \, \mathbb{T}  + \partial_x g\, R \otimes \mathds{1}  +\partial_x f\,\mathds{1}\otimes R^*\bigg)\right\}\, \\
    &\simeq \tr \prod_j \big(1+\varepsilon \mathbb{T}\big)  \big(\mathds{1} + \left[g(x_j+\varepsilon) - g(x_j)\right] R \otimes \mathds{1} \big)  \big(\mathds{1}+ \left[f(x_j+\varepsilon) - f(x_j)\right] \mathds{1} \otimes R^*\big) \label{eq:before_swap}\\
    &\simeq \tr \prod_j \big(1+\varepsilon \mathbb{T}+ g(x_j+\varepsilon)  [\mathbb{T},R\otimes \mathds{1}] + f(x_j+\varepsilon)  [\mathbb{T},\mathds{1} \otimes R^*]\big) \label{eq:after_swap}\, .
\end{align}
We go from line \eqref{eq:before_swap} to \eqref{eq:after_swap} by putting all the terms at point $x_j$ in the same factor of the product, using the commutation relations. Terms proportional to $R\otimes \mathds{1}$ and $\mathds{1}\otimes R^*$ cancel, and the remaining boundary terms vanish for test functions $f,g$ with compact support. Finally, noting that $[\mathbb{T},R\otimes \mathds{1} ]= [Q,R]\otimes\mathds{1}$, we have the result advertised in \eqref{eq:result_Y}.

Finally, to compute the expectation value itself, we use again a similar strategy the one used for monomials. 
We introduce $\rho(x):=\rho^{\beta_1 \beta_1}(x)$ with initial condition $\rho_{-\infty}=\rho_0$ and dynamics
\begin{equation}
    \frac{\upd}{\upd x} \rho(x) = \mathcal{L} \cdot \rho(x) + \beta_1 J(x) R \rho(x) +\beta_1 J(x) \rho(x) R^\dagger\, ,
\end{equation}
which is such that $\mathcal{Z}_{\beta_1J,\beta_1J} = \lim_{x\rightarrow+\infty} \tr[\rho(x)]$.

We then define the partial derivatives  $\rho^{(1,0)}:= \partial_{\beta_1} \rho|_{b_{1,2}=0}$, $\rho^{(0,1)}:= \partial_{\beta_1} \rho|_{b_{1,2}=0}$ and $\rho^{(1,1)}:= \partial_{\beta_1}\partial_{\beta_1} \rho|_{b_{1,2}=0}$. They obey the coupled ordinary differential equations
\begin{align}
    \frac{\upd}{\upd x} \rho^{(1,0)} &= \mathcal{L}\cdot \rho^{(1,0)}(x) + J(x) R \rho_0 \\
      \frac{\upd}{\upd x} \rho^{(0,1)}(x) &= \mathcal{L}\cdot \rho^{(0,1)}(x) + J(x)  \rho_0 R^\dagger \\
        \frac{\upd}{\upd x} \rho^{(1,1)} &= \mathcal{L}\cdot \rho^{(1,1)}(x) + J(x) R \rho(x)^{(0,1)} + J(x) \rho(x)^{(1,0)} R^\dagger \, .
\end{align}
Now we have that $\frac{\partial}{\partial \beta_1} \mathcal{Z}_{\beta_1J,\beta_1 J} = \lim_{x\rightarrow +\infty}\tr[\rho^{(1,1)}(x)$.
One can get an exactly analogous system of ODEs for a matrix $\sigma$ and its partial derivatives, replacing $R$ by $[Q,R]$ to get $\frac{\partial}{\partial \beta_1} \mathcal{Y}_{\beta_1J,\beta_1 J} = \lim_{x\rightarrow +\infty}\tr[\sigma^{(1,1)}(x)$. Finally, the expectation value we are looking for is obtained from the trace of the solutions
\begin{equation}
     \langle :h_\mathrm{fb}: \rangle = 2 \lim_{x\rightarrow +\infty}  \tr\left[m^2\rho^{(1,1)}(x) + \sigma^{(1,1)}(x)\right]\, .
\end{equation}
Again, the cost of evaluating this expectation value is $\propto D^3$.
\subsection{Strategy for other observables}
So far, we have obtained all the expectation values of local observables (one-point functions) that are needed for the optimization. We also obtained arbitrary (equal-time) $N$-point functions of $a(x)$. In general, one may also want $N$-point functions of more physically relevant observables like vertex operators. They can also all be obtained at a cost $\propto D^3$. The idea is again write the solution as (a derivative of) some generating functional which takes care of the convolution of $a(x)$ with $J$.

We can illustrate this idea on the two-point function 
\begin{equation}
    C_{\alpha_1,\alpha_2}(x) := \bra{Q,R} :\e^{\alpha_1\phi(x)} \!:\, :\e^{\alpha_2\phi(x)}\!:\ket{Q,R} \,.
\end{equation}
After expressing $\phi$ as a function of $a(x)$, one uses the BCH formula
\begin{align}
    C_{\alpha_1,\alpha_2}(x) :&= \bra{Q,R} \e^{\int \alpha_1J(y-x) a^\dagger(y)} \e^{\int \alpha_1J(y-x) a(y)} \, \e^{\int \alpha_2J(y) a^\dagger(y)} \e^{\int \alpha_2J(y) a(y)}\ket{Q,R} \\
    &=\e^{\int \alpha_1\alpha_2 J(y)J(x-y)}\bra{Q,R} \e^{\int \left[\alpha_1J(y-x) +\alpha_2 J(x)\right] a^\dagger(y)} \e^{\int \left[\alpha_1J(y-x) +\alpha_2 J(x)\right] }\ket{Q,R}\\
    &= \e^{\int \alpha_1\alpha_2 J(y)J(x-y)} \mathcal{Z}_{(\alpha_1J_x + \alpha_2J_0) , (\alpha_1J_x + \alpha_2J_0)}
\end{align}
with the notation $J_x(y) = J(x-y)$. Hence, computing this two-point function simply reduces to the evaluation of a one-point function but with a more complex source term. The cost is still $\propto D^3$. More general $N$-point functions of vertex operators $:\e^{\alpha_n\phi(x_n)}\!:$ can be obtained analogously, and one can get back to field monomials by differentiating with respect to $\alpha_n$.

\section{Geometric optimization of RCMPS}\label{app:geometric}
The optimization of RCMPS is non-trivial and relies heavily on Riemaniann techniques developed for CMPS \cite{vanderstraeten2019tangentspace,hauru2021}. To make the present paper self-contained, I explain the main technical ideas in this appendix. Most of this appendix thus does not describe new ideas, except for the very last part \ref{app:Riemann} that describes a slightly different regulation of the geometry than the standard one in \cite{hauru2021}.

\subsection{The RCMPS manifold}
Our objective is to see RCMPS as a Riemannian manifold, on which we can then use standard optimization techniques. The first step, following the standard CMPS approach, is to define tangent space vectors at a point $Q,R$ ~\cite{vanderstraeten2019tangentspace}
\begin{align}\label{eq:tangent}
    \ket{V,W}_{Q,R} =\!\! \int\! \upd x \left[V_{\alpha\beta} \frac{\delta }{\delta Q_{\alpha\beta}(x)} + W_{\alpha\beta} \frac{\delta }{\delta R_{\alpha\beta}(x)}\right] \! \ket{Q,R} \,
\end{align}
parameterized by two complex matrices $V,W$. Using the left canonical gauge $Q = -iK -\frac{1}{2} R^\dagger R $ as before, one can fix $V=-R^\dagger W$ without losing a linearly independent direction \cite{vanderstraeten2019tangentspace}. Hence there are only $2D^2$ independent real directions on the tangent space. Then, a natural choice of metric on this tangent space is to take the metric induced by the Hilbert space scalar product. One can show \cite{vanderstraeten2019tangentspace} that this overlap is simply
\begin{equation}\label{eq:overlap}
\begin{split}
    \langle W_1 | W_2 \rangle_{Q,R} &=  \tr [W_2\rho_0 W_1^\dagger]
    \end{split}
\end{equation}
where $\rho_\text{0}$ is the stationary state of the Lindbladian $\mathcal{L}$. The metric is then taken as simply the real part of this overlap \eqref{eq:overlap} 
\begin{equation}
    g(W_1,W_2)_{Q,R} = \text{Re}\left(\langle W_1 | W_2 \rangle_{Q,R}\right) \, .
\end{equation} 
Its computation is very fast, as it does not require solving an ODE, contrary to expectation values of local observables.

The next step is to understand how to move on the manifold to follow a descent direction, \ie\, we need to define a \emph{retraction}. It is simply a function $r$ that takes an initial point $(Q,R)$ on the manifold, a tangent space vector $W$, a parameter $\alpha$, and outputs a new point obtained by moving from $(Q,R)$ in the direction $W$ for a ``time'' $\alpha$. In principle, the natural retraction on a manifold is given by the geodesics. However, geodesics are typically expensive to compute, and it seems that flowing along them does not make optimization faster in general\footnote{I tried to optimize RCMPS using the geodesic retraction and approximate parallel transport. Computing the retraction then ends up dominating the optimization cost, without substantially reducing the number of iterations, at least on the examples I considered.}. Inspired by \cite{vanderstraeten2019tangentspace}, we can instead use the naive retraction
\begin{equation}
    (K,R)\rightarrow r(K,R,W,\alpha) = \bigg(K + \frac{i}{2}\left[W^\dagger(\alpha R+\alpha^2/2 W) -  (\alpha R^\dagger+\alpha^2/2 W^\dagger) W\right] \, , \,
    R + \alpha W \bigg)\, ,
\end{equation}
which does correspond to a move in the direction of $W$ and does preserve the gauge along the way ($K$ remains Hermitian).

\subsection{Gradients with backpropagation} \label{app:gradient}
Now that we know how to move on the RCMPS manifold, we need to know what is the direction that minimizes the energy. This is given by the gradient of the energy density, which we need to compute efficiently. This can fortunately be done with backward differentiation methods, for only twice the cost of computing the energy density itself. Let us illustrate the method on the expectation value of a vertex operator $\langle V_\beta\rangle_{Q,R}=\bra{Q,R} :\e^{\beta\phi}\!:\ket{Q,R}$ for $\beta$ real\footnote{For complex $\beta$, the expectation value $\langle V_\beta\rangle$ is not real, and thus $\nabla_{W} \langle V_\beta\rangle \neq g(\nabla \langle V_\beta\rangle,W) $. To get the gradient, one would first needs to split the expectation value into real and imaginary parts. Fortunately, in the Sine-Gordon model, only the real part of the vertex operators matter, and so we do not encounter this issue.}, but other local observables appearing in the Hamiltonian can be treated in the same way.

The gradient in the direction W, $\nabla_{W} \langle V_\beta\rangle = g(\nabla \langle V_\beta\rangle,W) $, is obtained via
\begin{equation}
     \langle V_\beta\rangle_{Q- \varepsilon R^\dagger W,R + \varepsilon W} =   V_\beta\rangle_{Q,R} + \varepsilon \nabla_{W} \langle V_\beta\rangle_{Q,R} + O(\varepsilon^2).
\end{equation}
Differentiating \eqref{eq:vertex_expectation} directly gives
\begin{equation}\label{eq:gradient_raw}
    \nabla_{W} \langle V_\beta\rangle=\int \upd y\, \tr\left\{ \left(\mathcal{P}\e^{\int_{y}^{+\infty} \mathcal{L}^\beta}\right) \cdot \nabla_{W}\mathcal{L}^\beta(y) \cdot\left(\mathcal{P}\e^{\int_{-\infty}^y \mathcal{L}^\beta}\right)\cdot \rho_\text{0}\right\}.
\end{equation}
with the notation $\mathcal{L}^\beta(x)\cdot \rho = \mathcal{L} \cdot \rho + \beta J(x) (R\rho + \rho R^\dagger)$ and
\begin{equation}\label{eq:derivative_Lindblad}
\begin{split}
    \nabla_{W}\mathcal{L}^\beta(y) \cdot \rho = -R^\dagger W \rho - \rho W^\dagger R+ \frac{1}{2} \left(R\rho W^\dagger+W\rho R^\dagger\right) 
    + \beta J(y) \left(W\rho + \rho W^\dagger\right)\, ,
    \end{split}
\end{equation}
We then replace the last part of the evolution from $y$ to $+\infty$ in \eqref{eq:gradient_raw} by the adjoint evolution applied to the identity
\begin{equation}\label{eq:gradient_adjoint}
\begin{split}
    \nabla_{W} \langle V_\beta\rangle=\int \upd y\, \tr\bigg\{ \left[\left(\mathcal{P}\e^{\int_{y}^{+\infty}  \mathcal{L}^{\beta*}}\right)\cdot \mathds{1}\right]  
    \times &\nabla_{W}\mathcal{L}^\beta(y)\cdot\left[\left(\mathcal{P}\e^{\int_{-\infty}^y \mathcal{L}^\beta}\right)\cdot \rho_\text{0}\right]\bigg\}.
\end{split}
\end{equation}
where the adjoint $\mathcal{L}^{\beta*}(y)$ of $\mathcal{L}^\beta(y)$ is defined as
\begin{equation}
    \mathcal{L}^{\beta *}(y)\cdot \mathcal{O} = Q^\dagger\mathcal{O} + \mathcal{O} Q+ \frac{1}{2} R^\dagger \mathcal{O} R + \beta J(y) \left[R^\dagger \mathcal{O} + \mathcal{O}R \right].
\end{equation}
The solution $\rho(x)=\mathcal{P}\e^{\int_{-\infty}^x \mathcal{L}^\beta}\!\cdot \rho_\text{0}$ of the forward problem and the solution  $\mathcal{O}_x=\mathcal{P}\e^{\int_{x}^{+\infty}  \mathcal{L}^{\beta *}}\!\cdot \mathds{1}$ of the backward problem can be computed by solving the corresponding ODEs. The gradient in the direction $W$ is then simply
\begin{align}
    \nabla_{W} \langle V_\beta\rangle &=\int \upd y \, \tr \left[\mathcal{O}_y \nabla_{W}\mathcal{L}^\beta (y)\cdot \rho_y\right] \\
    &=\tr\left[M_W W + M_{W}^\dagger W^\dagger\right]
\end{align}
with 
\begin{equation}\label{eq:matrices_integral} 
    M_W= \int \upd y\, -\rho_y \mathcal{O}_y R^\dagger + \frac{1}{2} \rho_y R^\dagger \mathcal{O}_y + \beta J(y) \rho_y \mathcal{O}_y 
\end{equation}
The matrix $M_W$ is obtained by evaluating the integrals with an efficient quadrature. Ultimately, the full gradient is obtained from the matrices M using $g(\nabla \langle V_\beta\rangle , W) = \nabla_W\langle V_\beta\rangle$ which gives
\begin{equation}
    \tr[ \nabla \langle V_\beta\rangle \rho_0 W + W^\dagger \rho_0 \nabla \langle V_\beta\rangle^\dagger ] = \tr[ M_W W + M_W^\dagger W^\dagger]
\end{equation}
and thus $\nabla \langle V_\beta\rangle = M_W \rho_0^{-1}$. 

The gradient of other observables can be computed with the same techniques. Hence, we have an efficient way to compute the gradient of the full energy density, and thus minimize it.

\subsection{Riemannian gradient descent and Quasi-Newton improvements} \label{app:Riemann}
Once we have a way to compute the gradient, and a way to move on the manifold in a given descent direction (a retraction), we can simply minimize the energy using gradient descent, that is descend in the direction opposite to the gradient. This is standard Riemannian gradient descent, which is equivalent to imaginary time TDVP in the tensor network literature. This already gives a fairly efficient method to minimize the energy, but becomes slower on ``difficult'' points, typically near phase transitions or for large values of $\beta$ in the Sinh-Gordon model.

In fact, we can pick even better descent directions using Riemannian quasi-Newton algorithms like Riemannian conjugate gradient or LBFGS. In addition to a retraction and a metric these algorithms require a way to transport vectors along a retraction path. The most natural  vector transport to use would be parallel transport, along with the geodesic retraction. Again, this is expensive to compute, and it turns out using the naive vector transport where tangent vectors stay the same along the retraction path gives almost as good results.

A final subtlety is related to the need for regulators and the use of a preconditioner. As was noted by \cite{hauru2021} in the context of MPS, the metric we use depends on $\rho_0$, the fixed point of $\mathcal{L}$, which can be very ill-conditioned. This is a problem since its inverse $\rho_0^{-1}$ appears in the gradient and damages the estimate of the Hessian made by LBFGS. One thus needs to regulate the metric by redefining the matrix $\rho_0 \rightarrow \rho_0^\varepsilon$ appearing in its definition $\rho^{\varepsilon}_0 = \rho_0 + \varepsilon \mathds{1}$. In practice, I noticed that a fixed regulator of order $10^{-2}- 10^{-3}$, \ie\, independent of the size of gradient norm, was needed for the LBFGS algorithm to get usable estimates of the Hessian, improving upon naive gradient descent. However, the descent part is much more efficient with a regulator that gets smaller as the gradient norm reduces as in \cite{hauru2021}. To combine both features, I used a Riemannian LBFGS with a strongly regulated metric defined with $\rho_0^\varepsilon$ for $\varepsilon=10^{-2}$, and added a preconditioner partially undoing the regulation as the gradient becomes small\footnote{The optimal prescription seemed to vary depending on the coupling, but a reliable choice was to target an effective metric with regulator $\varepsilon = 10^{-2} \| g(\nabla e_0,\nabla e_0)\|^2$.}. This strategy slightly differs from the one in \cite{hauru2021}, where the authors used the Euclidean metric (equivalent to $\varepsilon = +\infty$) for the LBFGS algorithm, and put the non-trivial regulated metric in the preconditioner only. Especially at large coupling and large bond dimensions, optimizing the state proved substantially faster by including the metric in the LBFGS algorithm as well.

\section{Implementation details}
The computations needed for this paper were carried in Julia, using an \emph{ad hoc} script leveraging many public packages. A near term objective will be to turn this inflexible piece of code into a legitimate public package adapted to generic QFT (at least perturbations of the bosonic free field). In the meantime, this appendix describes the strategy followed.

\subsection{Expectation values}
To evaluate the expectations values, I simply solved the ODEs mentioned in appendix \ref{app:expectations} for $x$ from $-20$ to $20$ (which provided results indistinguishable from $]-\infty, +\infty[$). The generator of the evolution has an integrable singularity in $x=0$ which I smoothed with the change of variable $u = \e^{-x^2} x^3+(1-\e^{-x^2}) x$. I used \texttt{KrylovKit.jl} to find the stationary state $\rho_0$ which is used in the initial condition. Then, to evolve the ODE forward, I used the package \texttt{DifferentialEquations.jl} \cite{rackauckas2017differentialequations} with the \texttt{Vern7} solver (an explicit Runge-Kutta scheme) and a relative tolerance of $10^{-12}$. The matrix multiplications necessary for the evaluation of the ODE generator were substantially sped up using \texttt{Octavian.jl}, which relies on \texttt{LoopVectorization.jl}.  For the bond dimensions I used (up to $D=32$), the only worthwhile parallelization was to compute the different expectation values appearing in the energy with different processes.

\subsection{Gradient}
To compute the gradient of the energy, I manually implemented the backpropagation technique presented in appendix \ref{app:gradient}. Doing so naively takes an excessive amount of memory, since the complete trajectory a priori needs to be stored to compute the integral \eqref{eq:matrices_integral}. However, one may only store points on the nodes of an efficient quadrature, much coarser than the grid used to solve the ODE. In practice, I found that using two tanh-sinh quadratures on $[-20,0]$ and $[0,20]$ provided machine precision estimates. Namely, taking a few hundred points was typically sufficient to make the quadrature error negligible compared to the ODE error.

\subsection{Optimization}

The energy minimization in this paper have been obtained with Riemannian LBFGS, with memory parameter $m=200$, using the remarkably convenient and problem agnostic package \texttt{OptimKit.jl}. All that is needed for the optimization is to provide the function to optimize (here the energy density), its gradient, the metric, a retraction, a vector transport, and an optional preconditioner that helps us revert part of the regulation as described in appendix \ref{app:Riemann}. The Riemannian conjugate gradient algorithm gave comparable performance for small coupling, but I observed that the LBFGS was faster for large coupling.

\bibliographystyle{apsrev4-1}
\bibliography{main}

\end{document}